\documentclass{aastex62}
\usepackage{natbib}

\received{*}
\revised{*}
\accepted{*}

\submitjournal{ApJ}
 
\shorttitle{ADIABATIC MASS LOSS IN BINARY STARS. III.}
\shortauthors{Ge et al.}

\begin{document}

\title{Adiabatic mass loss in binary stars. III. From the base of the red giant branch to the tip of asymptotic giant branch}
 
\correspondingauthor{Hongwei Ge; Ronald F Webbink}
\email{gehw@ynao.ac.cn; rwebbink@illinois.edu}

\author{Hongwei Ge}
\affiliation{Yunnan Observatories, Chinese Academy of Sciences, Kunming 650216, PR China}
\affiliation{Key Laboratory for the Structure and Evolution of Celestial Objects, Chinese Academy of Sciences, Kunming 650216, PR China}
\affiliation{Center for Astronomical Mega-Science, Chinese Academy of Sciences, Beijing 100012, PR China}
\affiliation{University of Chinese Academy of Sciences, Beijing 100049, PR China}

\author{Ronald F Webbink}
\affiliation{University of Illinois at Urbana-Champaign, 1002 W Green St, Urbana, 61801, USA}

\author{Xuefei Chen}
\affiliation{Yunnan Observatories, Chinese Academy of Sciences, Kunming 650216, PR China}
\affiliation{Key Laboratory for the Structure and Evolution of Celestial Objects, Chinese Academy of Sciences, Kunming 650216, PR China}
\affiliation{Center for Astronomical Mega-Science, Chinese Academy of Sciences, Beijing 100012, PR China}
\affiliation{University of Chinese Academy of Sciences, Beijing 100049, PR China}

\author{Zhanwen Han}
\affiliation{Yunnan Observatories, Chinese Academy of Sciences, Kunming 650216, PR China}
\affiliation{Key Laboratory for the Structure and Evolution of Celestial Objects, Chinese Academy of Sciences, Kunming 650216, PR China}
\affiliation{Center for Astronomical Mega-Science, Chinese Academy of Sciences, Beijing 100012, PR China}
\affiliation{University of Chinese Academy of Sciences, Beijing 100049, PR China}

\begin{abstract}

The distinguishing feature of the evolution of close binary stars is the role played by the mass exchange between the component stars. Whether the mass transfer is dynamically stable is one of the essential questions in binary evolution. In the limit of extremely rapid mass transfer, the response of a donor star in an interacting binary becomes asymptotically one of adiabatic expansion. We use the adiabatic mass loss model to systematically survey the thresholds for dynamical timescale mass transfer over the entire span of possible donor star evolutionary states. We also simulate mass loss process with isentropic envelopes, the specific entropy of which is fixed to be that at the base of the convective envelope, to artificially mimic the effect of such mass loss in superadiabatic surface convection regions, where the adiabatic approximation fails. We illustrate the general adiabatic response of $3.2 M_{\odot}$ donor stars at different evolutionary stages. We extend our study to a grid of donor stars with different masses (from 0.1 to 100 $M_{\odot}$ with Z = 0.02) and at different evolutionary stages. We proceed to present our criteria for dynamically unstable mass transfer in both tabular and graphical forms. For red giant branch and asymptotic giant branch donors in systems with such mass ratios, they may have convective envelopes deep enough to evolve into common envelopes on a thermal timescale, if the donor star overfills its outer Lagrangian radius. Our results show that the red giant branch and asymptotic giant branch stars tend to be more stable than previously believed, and this may be helpful to explain the abundance of observed post-AGB binary stars with an orbital period of around 1000 days.

\end{abstract}


\section{INTRODUCTION}
\label{intro}

The distinguishing feature of the evolution of close binary stars is the role played by the mass exchange between the component stars. Whereas the evolution of a single star is dictated primarily by its initial mass, and, to a lesser extent, its heavy element abundances, the possibility of mass exchange in a binary star introduces at least two additional dimensions needed to describe their initial configurations, namely the initial mass of the companion star and the orbital period (or orbital semi-major axis) of the binary. These additional dimensions give rise to myriad evolutionary channels and observable configurations unique to binary stars.

A very large fraction of all stars are members of binary or multiple systems \citep[e.g.,][]{moe17}. Binary evolution is related to almost all classes of objects and phenomena, such as type Ia supernovae \citep{wang12,meng17}, double neutron stars/black holes \citep{abbo17a,abbo17b}, barium stars and other chemically polluted stars \citep{boff88}, blue stragglers \citep{math09}, planetary nebula nuclei \citep{jone17}, red novae \citep{kami15}, symbiotic stars \citep{soko17}, super-soft X-ray sources \citep{vand92,kaha97}, and references therein. Understanding binary evolution, and the physical processes which govern it, is therefore an essential element in building a complete picture of the stellar content of the local universe. It is toward that goal that this series of papers is directed.

A crucial parameter in determining which evolutionary channel a particular binary will follow is the timescale on which mass transfer proceeds once one of the stellar components fills its Roche lobe. As described in Paper I of this series \citep{ge10}, three natural timescales arise in this context. At one extreme, mass transfer is driven solely by a combination of the evolutionary expansion of the donor star and orbital contraction due to angular momentum loss. This is generally the slowest mode of mass transfer, and we will refer to it as \emph{evolutionary mass transfer}. Mass transfer can also be driven on the thermal timescale of the donor star. This situation arises when the donor is driven out of thermal equilibrium by mass loss, but contracts in parallel with its Roche lobe. This \emph{thermal timescale mass transfer} is self-regulating. Perturbation of the mass transfer rate triggers a negative feedback: increasing/decreasing the mass transfer rate leads to the donor contracting more/less rapidly than its Roche lobe. However, at mass loss rates greatly exceeding thermal timescale mass loss, thermal relaxation in the donor interior cannot compete with decompression due to mass loss. In the asymptotic limit of arbitrarily high mass loss rates, the donor interior responds adiabatically to decompression, and if the donor is then still unable to contract as rapidly as its Roche lobe, the mass transfer rate is subject to positive feedback. This \emph{dynamical timescale mass transfer} rate is then limited only by the mass of the donor envelope and the orbital period.

Paper I of this series described the construction of adiabatic stellar mass loss sequences. In these sequences, the specific entropy profile and composition profile of the donor star are held fixed during mass loss. These sequences then describe the asymptotic response of the donor to mass loss so rapid that its thermal relaxation can be neglected, but not so rapid that the donor departs in bulk from hydrostatic equilibrium \citep{pacz72,savo78,eggl06}.

As described in Paper I, beginning with a stellar evolutionary model, we construct a model sequence of decreasing stellar mass but fix composition and specific entropy profiles. This adiabatic mass-radius relation ($\zeta_{\rm ad}$) is then compared with a family of Roche lobe mass-radius relations ($\zeta_{\rm L}$) sharing the same initial mass and radius, but with a range of possible mass ratios (donor/accretor). We assume for simplicity that mass transfer is conservative of mass and orbital angular momentum. In this simplest approximation, the critical mass ratio ($q_{\rm cr}$) for the onset of dynamical timescale mass transfer is that for which the adiabatic mass-radius relation just reaches, but does not exceed the Roche lobe mass-radius relation.

In this paper, we complete the survey of threshold conditions for dynamical timescale mass transfer which began in Paper II of this series \citep{ge15}. That paper concerned donor stars with radiative envelopes; the present paper presents results for donors with convective envelopes, while updating the results from Paper II. We characterize thresholds for dynamical timescale mass transfer in terms of a critical mass-radius exponent $\zeta_{\rm ad} = \zeta_{\rm KH} \equiv {\rm d}\ln R_{\rm KH}/ {\rm d}\ln M = \zeta_{\rm L}$. As we proposed in Papers I and II, we compare the donor star's inner radius ($R_{\rm KH}$) response with its Roche lobe radius ($R_{\rm L}$) response to obtain the mass ratio thresholds for dynamical timescale mass transfer, $q_{\rm ad}$. The matter between the inner radius, $R_{\rm KH}$, and surface radius, $R$, is assumed to be an adiabatic flow, which transfers to its companion through the inner Lagrangian point, ${\rm L_1}$, and can be accelerated to a thermal timescale mass transfer rate, $\dot{M}_{\rm KH} = -M_{\rm i}/\tau^{\rm i}_{\rm KH}$. The subscript and superscript, ${\rm i}$, indicate initial values when the donor overfills its Roche lobe.

To provide a framework for understanding the full scope of our results, we offer in Section 2 an account of the adiabatic response to mass loss of a $3.2\ M_{\odot}$ star as it evolves from the zero-age main sequence (ZAMS) through the main sequence (MS) and the Hertzsprung gap (HG), to the tip of the red giant branch (TRGB), through core helium ignition, up to the onset of core carbon burning , or the tip of the asymptotic giant branch (TAGB). This account will illustrate the abrupt change in the character of that response as this star transitions from a radiative envelope to a convective one as it evolves through the base of the giant branch (BGB). As this star then evolves first up the red giant branch (RGB) and then up the asymptotic giant branch (AGB), the dramatic impact of superadiabatic convection becomes apparent. In Section 3, then, we describe our treatment of mass loss from superadiabatic surface layers and address the limitations of one-dimensional models in addressing this phenomenon. Section 4 then presents the full range of results of our adiabatic mass loss calculations. Section 5 compares these results with prior attempts to delineate the threshold conditions for dynamical timescale mass transfer. We emphasize the importance of incorporating physically and dimensionally appropriate models for mass outflow from the inner critical Roche point to couple the mass transfer rate to the radial extent by which the donor overfills its Roche lobe. We conclude in Section 6  with a discussion of the limitations of our models and the need for thermal-equilibrium mass loss models.

\section{ADIABATIC MASS LOSS: A CASE STUDY}

Consider a typical, intermediate-mass ($3.2~M_{\odot}$) star as it evolves from the ZAMS to core carbon ignition. Its observable progress is most often portrayed in terms of luminosity and effective temperature in a Hertzsprung-Russell diagram. Figure 1 shows the evolutionary track followed by our $3.2~M_{\odot}$ example. Major mileposts are labeled thereon. In the context of binary star evolution, however, it is the radius of a star as it first fills its Roche lobe that is of prime importance, since it is this event that marks the onset of tidal mass transfer, and it is the mean density of the lobe-filling star that fixes the binary orbital period (to within a weak function of the binary mass ratio). Figure 2 shows the radius of our $3.2\ M_{\odot}$ star as a function of time. Again, major mileposts are labeled. This figure also identifies those evolutionary phases that are "shadowed", that is, those phases in which the star's radius is smaller than in a preceding phase, and should have triggered tidal mass transfer in that preceding phase. In our $3.2\ M_{\odot}$ example, core helium burning, from ignition through central helium exhaustion, is such a shadowed phase.

\begin{figure}[ht!]
	\centering
	\includegraphics[scale=0.36]{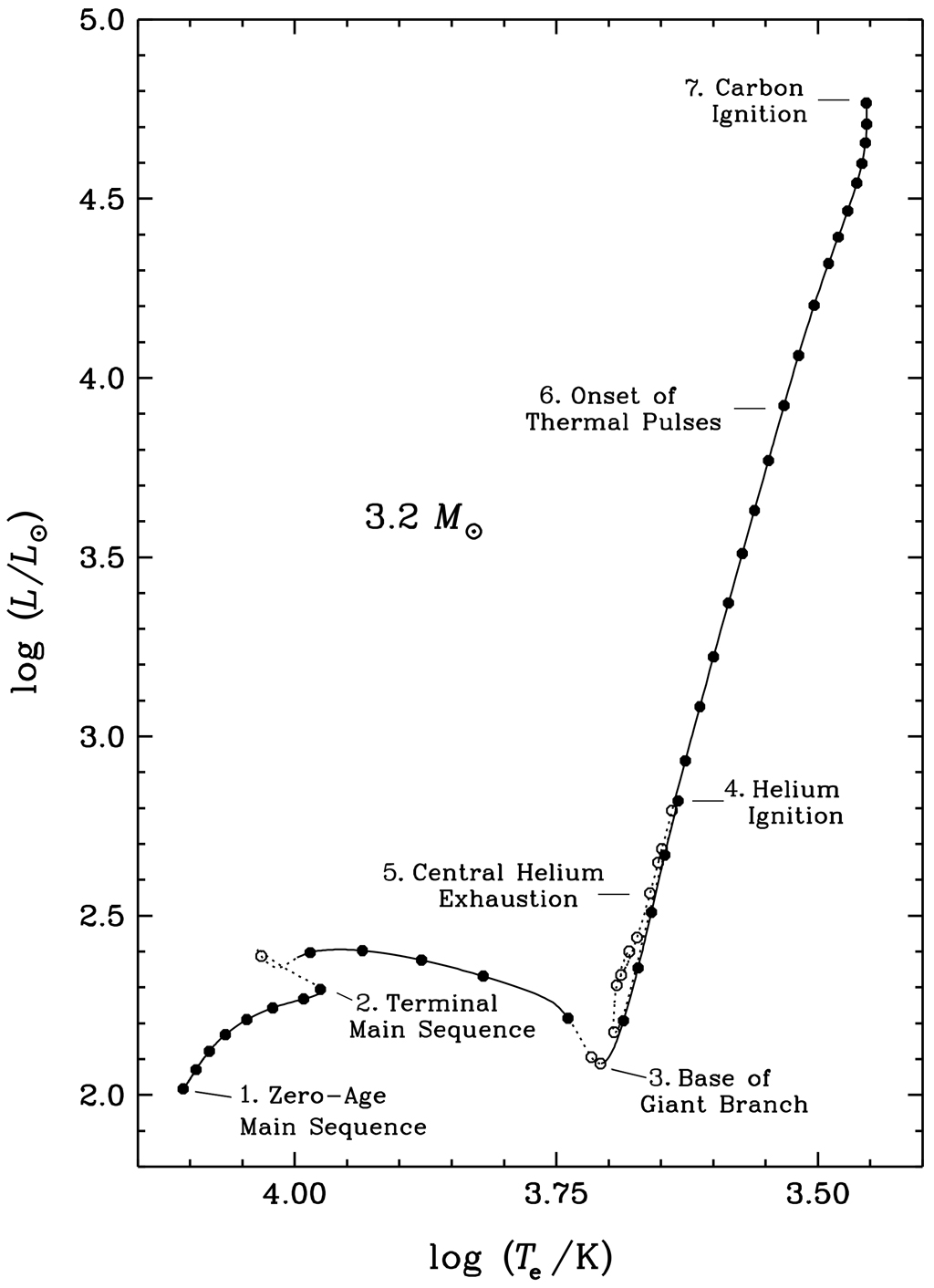}
	\caption{The evolutionary track of a $3.2 M_{\odot}$ star in the Hertzsprung-Russel diagram. The models we choose to study in its evolutionary track are labeled with solid dots and circles. The thin dotted line of the evolutionary track signifies evolutionary phases in which the stellar radius is smaller than in the preceding phase. Seven major evolutionary phases are labeled. \label{3_2HRD}}
\end{figure}

\begin{figure}[ht!]
	\centering
	\includegraphics[scale=0.36]{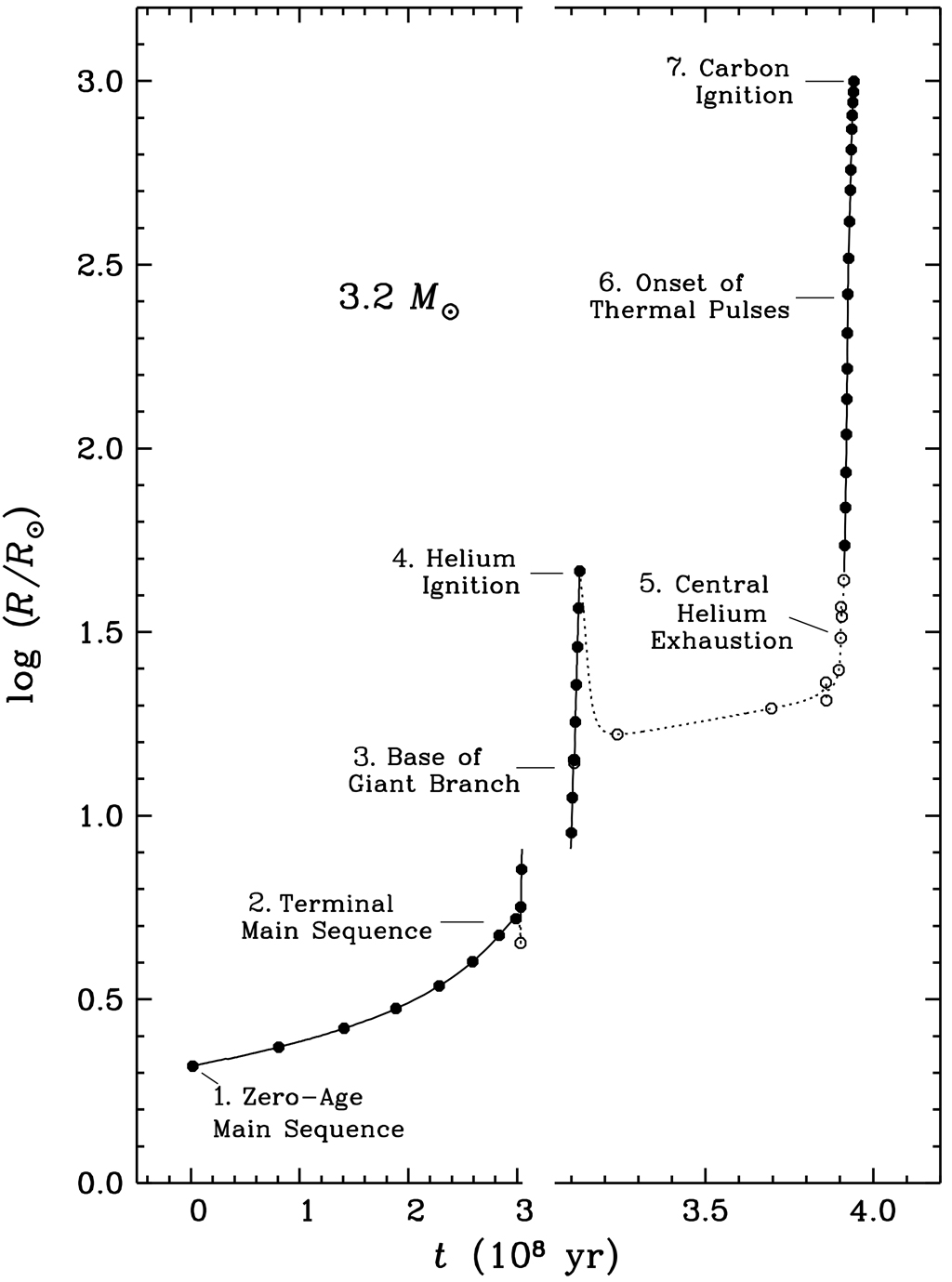}
	\caption{The radius of a $3.2 M_{\odot}$ star as a function of its age. The stellar models we study in its evolutionary track are labeled with solid dots and circles. Seven major evolutionary phases, which are the same as shown in Figure~\ref{3_2HRD}, are labeled here. Absent significant angular momentum loss, this star cannot initiate mass transfer during those phases of its evolution when its radius is smaller than during a preceding phase of evolution (thin dotted line, e.g., during core helium burning, or just beyond the point labeled terminal main sequence (TMS)). We refer to these forbidden evolutionary phases as "shadowed". For reviewing clearly the rapid evolution in advantage phases, a break is made in the horizontal axis and the axis spacing is reduced by half. \label{3_2RtD}}
\end{figure}

The evolution of our stellar example reflected in Figures~\ref{3_2HRD} and~\ref{3_2RtD} is intimately connected to its interior composition and thermal structure. Inasmuch as we are focused here on the asymptotic (adiabatic) response of a star to mass loss, it is most useful to characterize the thermal structure in terms of the specific entropy profile of the mass-losing star. This profile, and that of the nuclear composition, are fixed along a given adiabatic mass loss sequence.

\begin{figure}[ht!]
	\centering
	\includegraphics[scale=0.36]{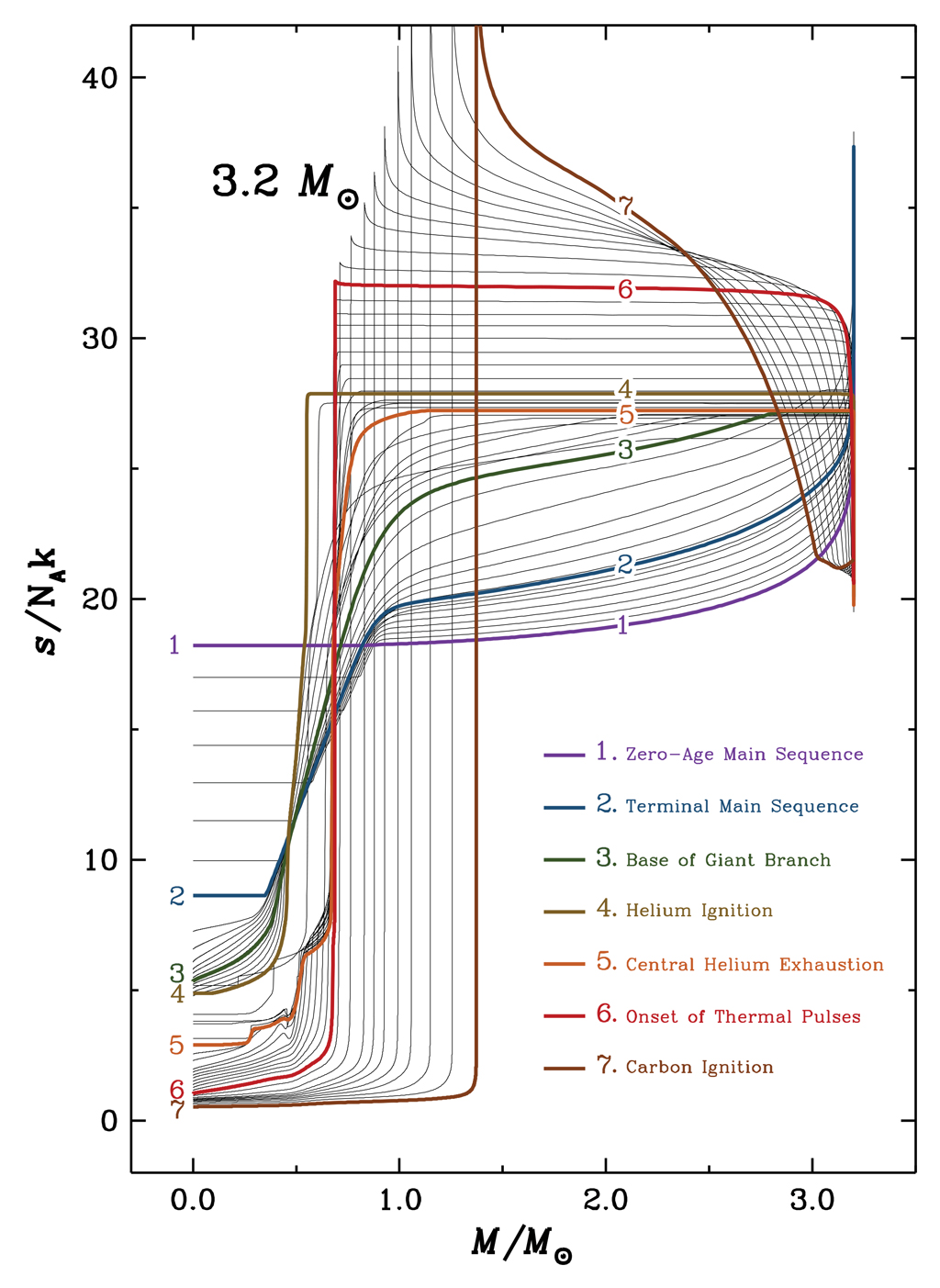}
	\caption{Specific entropy profiles of a $3.2 M_{\odot}$ star that we study in its whole evolutionary track. Thick colored lines indicate the entropy profile from the stellar models that are labeled in Figures~\ref{3_2HRD} and~\ref{3_2RtD}.\label{3_2sMD}}
\end{figure}

The evolution of the specific entropy profile of our $3.2\ M_{\odot}$ star is portrayed in Figure~\ref{3_2sMD}. Bearing in mind the (Schwarzschild) criterion for stability against convection (that specific entropy increase outwards), we see here the existence of a contracting hydrogen-depleted convective core during MS evolution (flat portion of the profile over the central 0.8 - 0.3 $M_{\odot}$). Outside the core, specific entroy increases with mass or radius, so the envelope is strongly stable against convection. As the core reaches hydrogen exhaustion, its contraction accelerates (the core is not yet electron-degenerate in this example), and energy generated by that contraction is largely absorbed in the surrounding envelope. The specific entropy of the envelope thus increases rapidly, driving rapid envelope expansion as the star crosses the Hertzsprung gap (see Figures~\ref{3_2HRD} and~\ref{3_2RtD}). That expansion is arrested when the stellar photosphere becomes so cool and tenuous that it begins to turn transparent. This cutoff of opacity (the Hayashi limit) marks the star's arrival at the base of its giant branch evolution. Up to this point, our $3.2\ M_{\odot}$ star has maintained a radiative outer envelope, but it now develops a deepening surface convection zone (note the development of an extended flat entropy profile in Figure~\ref{3_2sMD}) that it will support for the duration of its evolution. Meanwhile, deep in its core compression leads to helium ignition. With the onset of core helium burning, hydrogen burning in the surrounding shell subsides, leading to a contraction in total radius and a decline in stellar luminosity. As central helium approaches exhaustion, the site of helium-burning shifts to a surrounding shell and advances outwards toward the largely dormant hydrogen-burning shell. The star begins its ascent of the AGB. The inert carbon-oxygen central core left behind by helium-burning contracts, and becomes electron-degenerate. The helium-burning shell is drawn deeper into the gravitational potential of this core, burning through its surrounding helium layer, until its propagation outward in mass is constrained by the rate of advancement of the hydrogen-burning shell. As the two burning shells converge in mass, hydrogen-burning is stimulated, while the growth of the helium-burning luminosity momentarily hesitates, before becoming thermally unstable. This marks the beginning of the thermally-pulsing asymptotic giant branch (TPAGB) phase characterizing the remainder of the star's evolution. The combination of very high luminosity and very extended stellar envelope make convection very inefficient as reflected in the very pronounced negative (superadiabatic) specific entropy profile that develops during this phase (Figure~\ref{3_2sMD}).

\begin{figure}[ht!]
	\centering
	\includegraphics[scale=0.36]{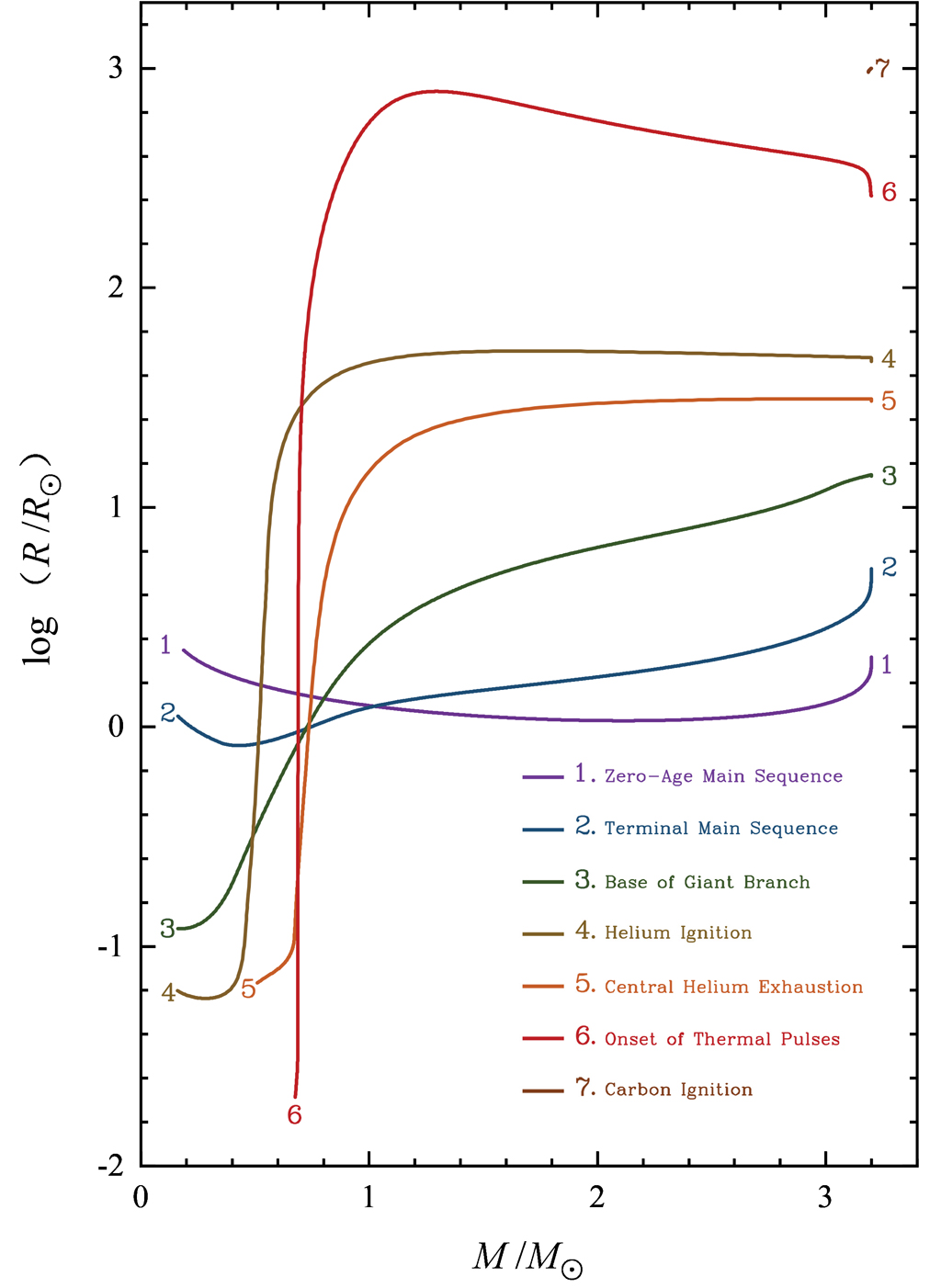}
	\caption{The radius responses to adiabatic mass loss of a $3.2\ M_{\odot}$ star that on different evolutionary stages. The initial stellar models are derived from standard evolutionary sequences with a mixing-length convection envelope. Colored lines represent the radius response of the stellar models that are labeled in Figures~\ref{3_2HRD},~\ref{3_2RtD}, and~\ref{3_2sMD}. We can notice that the last stellar model terminates unexpectedly, and we address the physical explanation in the last section. \label{3_2RM-d}}
\end{figure}

The connection between stellar specific entropy profile and the characteristic adiabatic response of a star to mass loss --- whether expansion or contraction --- can now be illustrated with reference to Figures~\ref{3_2sMD} and~\ref{3_2RM-d}. Prior to its arrival at the BGB, our $3.2 M_{\odot}$ star has a radiative envelope. Specific entropy increases rapidly toward the surface. When the matter is lost from the surface of this star, lower-entropy gas --- that is, cooler and denser gas --- is exposed, leading to a rapid contraction in photospheric radius. The adiabatic responses of stars with radiative envelopes were documented in Paper II of this series.

In the case of stars with convective envelopes, on the other hand, mass loss from the surface exposes underlying gas of the same (or greater) specific entropy. The responses of stars, in this case, is mediated by a gradual weakening in surface gravity, allowing the star to expand in radius, unless the surface convection zone is relatively shallow. If surface convection is inefficient (superadiabatic), gas brought to the surface by mass loss is with a {\it higher} specific entropy, that is, hotter and more rarified. The photospheric radius then expands quite rapidly, as the stellar envelope is no longer contained by a denser, low-entropy surface layer. For stars on the AGB, this expansion can be quite spectacular, as we shall discuss in the next section.

\begin{figure}[ht!]
\centering
\includegraphics[scale=0.50]{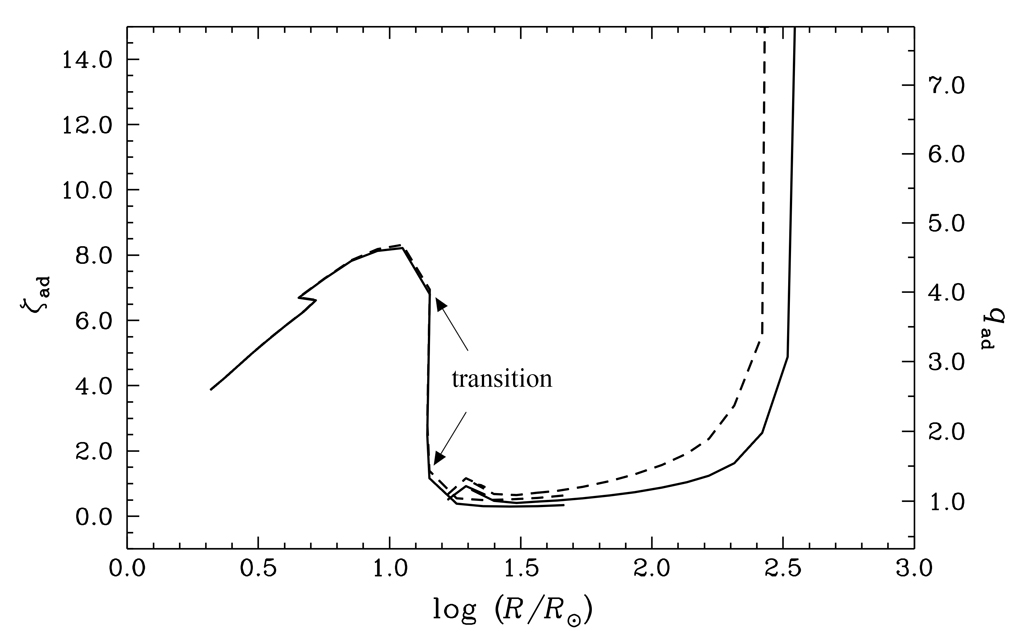}
\caption{The critical mass-radius exponent $\zeta_{\rm ad}$ and mass ratio $q_{\rm ad}$ as functions of stellar radius for the $3.2\ M_{\odot}$ models. The arrows point out the abrupt transition from delayed dynamical instability to prompt dynamical instability at ${\rm log} (R/R_\odot)=1.15$ ($R = 14 R_\odot$), as the star approaches the base of the giant branch (BGB). The solid curve corresponds to models with standard mixing-length envelopes ($\zeta_{\rm ad}$ and $q_{\rm ad}$), and the dashed curve to models with artificially isentropic convective envelopes ( $\widetilde{\zeta}_{\rm ad}$ and $\widetilde{q}_{\rm ad}$). \label{3_2zeta}}
\end{figure}

With the physical explanation of the stellar responses to adiabatic mass loss, it would be easy to understand how the critical mass ratio $q_{\rm ad}$ changes for stars at different evolutionary stages. Before we deal with the mass transfer instability of $3.2\ M_{\odot}$ stars, we address here in more detail how we solve for the critical mass ratio, $q_{\rm ad}$. It would be very simple for us to calculate $q_{\rm ad}$ if the stars' mass-radius exponent $\zeta_{\rm ad}$ equals $\zeta_{\rm L}$. However, the onset of dynamical timescale mass transfer cannot be instantaneous, but rather accelerates from an initial trickle to full-blown dynamical instability as it surpasses the stellar thermal timescale rate, that is, as the flow asymptotically becomes adiabatic. Therefore, we define the onset of dynamical timescale mass transfer not from the instant a donor star fills its Roche lobe, but rather from the instant at which the Roche lobe penetrates deeply enough into the stellar envelope to drive thermal timescale mass transfer, $\dot{M} = \dot{M}_{\rm KH}$\footnote{$\dot{M}_{\rm KH}= R_{\rm i} L_{\rm i}/G M_{\rm i}$, and $\dot{M}$ is integrated from interior radius $R_{\rm KH}$, to its surface radius, $R$, see formula A(9) in Paper I.}. Thus, the mass-radius exponent, $\zeta_{\rm ad} \equiv \zeta_{\rm KH}$, actually derives from how the interior radius, $R_{\rm KH}$, responds to its mass $M$. With this definition of the mass-radius exponent in mind, we hereby discuss how $\zeta_{\rm ad}$ and $q_{\rm ad}$ vary at different evolutionary stages. The adiabatic mass-radius exponents, $\zeta_{\rm ad}$, and critical mass ratios, $q_{\rm ad}$, increase gradually with the increase of the initial radii of the $3.2 M_{\odot}$ MS/HG stars (the solid line in Figure~\ref{3_2zeta} with ${\rm log} (R/R_\odot) < 1.15$). This means that, as their convective cores shrink and radiative envelopes expand, $3.2\ M_{\odot}$ MS/HG stars become more stable against dynamical timescale mass transfer. Similar to the $5\ M_{\odot}$ stars in Paper II, when the mass of the convective envelope ($M_{\rm ce}$) reaches approximately $10^{-3} M_{\rm i}$, the critical conditions for dynamical timescale mass transfer undergo an abrupt, but continuous, transition from delayed to prompt instability (where ${\rm log} (R/R_\odot) = 1.15$ in Figure~\ref{3_2zeta}). 

\section{SUPERADIABATIC MASS LOSS}
\label{sec_supad}

As described in Paper I, the construction of adiabatic mass loss sequences offers the attraction of estimating threshold conditions for dynamical timescale mass transfer, without the need to resort to time-dependent models to search for those thresholds. This approach is possible because, even at very high mass loss rates, the breakdown of hydrostatic equilibrium entailed in driving mass outflow from the donor star is concentrated in the neighborhood of the inner Lagrangian point, $L_1$. Far from $L_1$, the mass flow is subsonic \citep{pacz72,savo78,eggl06} and the departure from local hydrostatic equilibrium is negligibly small. The limiting dynamical response to mass loss then depends only on the maintenance of hydrostatic equilibrium through the bulk of the donor interior. If the mass loss is rapid enough, energy transport cannot keep pace, and relaxation to hydrostatic equilibrium becomes asymptotically adiabatic.

In reality, this approximation cannot hold throughout the donor. Sufficiently near the donor photosphere, the energy density in gas falls below that in radiation, and radiative relaxation becomes much faster than the local dynamical timescale. That is, even for arbitrarily rapid mass transfer, there will always exist a region just below the stellar photosphere where the adiabatic approximation fails. If this region in the donor star is in radiative equilibrium, this breakdown in the adiabatic approximation is of little consequence, as the decompression of this region is typically accompanied by a rapid rise in opacity that chokes off radiative outflow (cf. Figure 7 of Paper II). But in a superadiabatic convective envelope, the adiabatic loss of the surface layers would bring higher entropy gas to the surface, prompting an abrupt expansion that stimulates even faster mass loss. This positive feedback is inherently destabilizing.

Since the focus of this paper is on estimating threshold conditions for the onset of dynamical timescale mass transfer, we wish to quantify the possible impact of superadiabatic mass loss on those thresholds. To that end, we construct mass loss sequences that parallel those of donor stars with conventional mixing-length convective envelopes, but in which those convective envelopes have been replaced by isentropic envelopes, with specific entropy fixed to be that at the base of the convective envelope (see Section3.4 in Paper I). That is, we artificially suppress the decline in specific entropy with radius in the outer envelope that characterizes superadiabatic convection. So constructed, these artificially isentropic envelope models mimic the behavior of time-dependent mass loss models, in which the outer entropy profiles migrate homologously inward with mass loss (see, for example, Figure 2 in \citealt{wood11}). For that reason, we consider the threshold mass-radius exponent and corresponding critical mass ratio derived from the isentropic envelope models to be more reflective of reality than are the strictly adiabatic mass loss sequences. The rationale and construction of isentropic envelope models are described in greater detail in Papers I and II.

As discussed in the last section, even in systems which become unstable due to dynamical timescale mass transfer, that instability does not manifest itself immediately at the onset of mass transfer, but, rather, the mass loss rate from the donor accelerates until it surpasses the thermal timescale rate, $|\dot{M}| > M/\tau_{\rm KH}$, where $\tau_{\rm KH} = GM^2/RL$ is the nominal thermal (Kelvin-Helmholtz) timescale of the donor.  Beyond that point, the mass transfer becomes fast enough to freeze the donor entropy profile progressively from its deep interior to its surface. We, therefore, associate the critical mass ratio $q_{\rm ad}$ (for conservative mass transfer) or, more generally, the critical adiabatic mass-radius exponent, $\zeta_{\rm ad} \equiv (\partial \ln R/\partial \ln M)_{\rm KH}$, with the condition that the donor mass loss rate only just manages to reach $M/\tau_{\rm KH}$.  We label the mass and (interior) Roche lobe radius of the donor at this critical point $M_{\rm KH}$ and $R_{\rm KH}$, respectively; the surface (photospheric) radius of the donor at this point is $R^*_{\rm KH}$.  That is, the donor overfills its Roche lobe by an amount $R^*_{\rm KH} - R_{\rm KH}$ just sufficient to drive thermal timescale mass transfer. The corresponding parameters for the models with isentropic envelopes are denoted by a superscript tilde, e.g., $\tilde{M}_{\rm KH}$, $\tilde{R}_{\rm KH}$, $\tilde{R}^*_{\rm KH}$, $\tilde{\zeta}_{\rm ad}$, and $\tilde{q}_{\rm ad}$. Because the isentropic-envelope models have higher-entropy envelopes than do those models with standard mixing-length envelopes, they have larger radii. We characterize this radius excess by the parameter $\Delta_{\rm exp} \equiv \log (\tilde{R}_{\rm i}/R_{\rm i})$, where $\tilde{R}_{\rm i}$ and $R_{\rm i}$ are the initial radii of donors with isentropic and mixing-length envelopes, respectively, at the onset of mass transfer (see Figure 6 of Paper I).

Despite the appearance that this model of superadiabatic mass loss is physically unrealistic, its behaviour is not without a physical basis. The models computed here are one-dimensional approximations to a three-dimensional configuration. Mass outflow from the donor star is far from spherically symmetric but rather focused through a gravitational nozzle toward the companion star. In a radiative star, we expect that mass outflow to be fueled primarily by surface flows in the donor envelope since its envelope is stably stratified against buoyancy forces. In a convective star, in contrast, we expect the mass outflow to be driven primarily by buoyancy forces along the line of centers. Rising buoyant convective elements need not be balanced by denser descending elements, but maybe preferentially captured by the binary companion. Further complications are introduced by Coriolis effects on the outflow, which will tend to break it up into turbulent eddies, and of course, magnetic fields could be dynamically important. Satisfactory resolution of the impact of these competing phenomena calls for full three-dimensional models, including radiative transport, convection, turbulence, and magnetic fields, all in a rotating reference frame --- a capability well beyond current computational capabilities.

Now we come back to the thresholds for the dynamical timescale mass transfer of the RGB/AGB donor stars (Figure~\ref{3_2zeta} with ${\rm log} (R/R_\odot) > 1.15$). The adiabatic mass-radius exponents, $\zeta_{\rm ad}$, and critical mass ratios, $q_{\rm ad}$, increase gradually with the initial radii of the $3.2 M_{\odot}$ AGB stars (the solid line in Figure~\ref{3_2zeta} with ${\rm log} (R/R_\odot) > 1.7$). Although the size of the convective envelope increases as the radius of the host $3.2 M_{\odot}$ AGB star increases, the envelope becomes more diffuse. So it becomes harder to reach a thermal timescale mass transfer, $\dot{M} = \dot{M}_{\rm KH}$ as the radius difference between $R^*_{\rm KH}$ and $R_{\rm KH}$ increases if a $3.2 M_{\odot}$ AGB star on a late evolutionary stage (${\rm log} (R/R_\odot) > 2.5$). Therefore, the critical mass ratio, $q_{\rm ad}$, also increases dramatically. However, the radius difference between $R^*_{\rm KH}$ and $R_{\rm KH}$ is smaller for RGB donors than for AGB donors. Hence, the competition between the increase of the mass of the convective envelope and that of $R^*_{\rm KH} - R_{\rm KH}$ leads to the $3.2 M_{\odot}$ RGB stars having a nearly constant adiabatic mass-radius exponent, $\zeta_{\rm ad}$, or critical mass ratio, $q_{\rm ad}$ (Figure~\ref{3_2zeta}). For the stellar models with isentropic convection envelopes, the adiabatic expansion is supressed slightly. Thus, $\tilde{\zeta}_{\rm ad}$ and $\tilde{q}_{\rm ad}$ for the stellar models with an isentropic envelope are larger than their standard mixing length envelope counterparts.

\section{RESULTS}
\label{sec_res}

In Paper II, we surveyed the adiabatic responses of Population I ($Z = 0.02$) stars spanning a full range of stellar mass ($0.10\ M_{\odot}$ to $100\ M_{\odot}$) and evolutionary stages from the ZAMS to the terminal main sequence (TMS), through the HG to the BGB. In the present paper, we update and extend our survey of adiabatic mass loss to late phases of stellar evolution, from the ZAMS through the MS, HG, RGB, and AGB, up to the exhaustion of the hydrogen-rich envelopes, carbon ignition, or core-collapse, as the case may be. Our initial stellar model sequences were updated as follows: (1) We enhance the spatial resolution of our code to deal with more complex structures, i.e., we increased the number of the mesh points to 1000 for RGB and AGB simulations; (2) We improve our prescription for overshooting, and thereby increase the mixing efficiency for intermediate-mass and massive stars, resulting in a smoother composition profile near the stellar core. We select the stellar models using the same method in Section 3 of Paper II, but the model selection is revised accordingly.

For a better understanding of our results, we briefly summarize the general response of the giant branch donor stars (low- and intermediate-mass) to adiabatic mass loss. In these evolutionary phases, all RGB and AGB stars develop deep convective envelopes, and so react very differently to mass loss than do stars in the MS and HG that were the focus of Paper II. The onset of convection is (absent composition gradients) triggered by a negative specific entropy gradient, i.e., specific entropy decreasing with increasing radius. Where convection is very efficient, carrying the bulk of the stellar luminosity, that gradient is typically very small. On the other hand, where convection is inefficient, typically in tenuous outer envelopes, especially of very luminous stars, the local temperature gradient becomes markedly \emph{superadiabatic}, that is, steeper than the local adiabatic temperature gradient. In simplest terms, then, gas lost from the surface of a convective envelope is replaced by a gas of equal or greater specific entropy. The reduction in local gravity due to mass loss may then result in a net expansion in the stellar radius, which can be quite dramatic if the surface layers are markedly superadiabatic. This is a response clearly conducive to dynamical instability. We now present our results quantitatively as follows.

Tables \ref{intmod} and \ref{glbmod} document the initial properties of the donor stars at the beginning of each adiabatic mass-loss sequence.

\begin{deluxetable}{rrrrrrrrrrr}   
     \tabletypesize{\footnotesize}
     \tablewidth{0pt}
     \tablecolumns{11}
     \tablecaption{Interior properties of initial models\label{intmod}}
     
     \tablehead{
          \colhead{$k$} & \colhead{$\log t$} & \colhead{$M_{\rm ce}$} & \colhead{$M_{\rm c}$} &
          \colhead{$M_{\rm ic}$} & \colhead{$\psi_{\rm c}$} & \colhead{$\log \rho_{\rm c}$}
&
          \colhead{$\log
               T_{\rm c}$} & \colhead{$X_{\rm c}$} & \colhead{$Y_{\rm c}$} &
\colhead{$X_{\rm s}$} \\
          \colhead{} & \colhead{yr} & \colhead{$M_\sun$} & \colhead{$M_\sun$} &
\colhead{$M_\sun$} &
          \colhead{} &
          \colhead{$\rm g\:cm^{-3}$} & \colhead{K} & \colhead{} & \colhead{} & \colhead{}
     }
     \startdata
     \cutinhead{\normalsize $3.2000\ M_\sun$}
     1  & 6.208016 & 0.0000 & 0.8850 & 0.0000 & --3.465 & 1.546 & 7.383 & 0.698 & 0.282 &
0.700 \\*
     2  & 7.907971 & 0.0000 & 0.8474 & 0.0000 & --3.580 & 1.528 & 7.386 & 0.599 & 0.381 &
0.700 \\
     3  & 8.149788 & 0.0000 & 0.8102 & 0.0000 & --3.677 & 1.525 & 7.394 & 0.500 & 0.480 &
0.700 \\
     4  & 8.276208 & 0.0000 & 0.7728 & 0.0000 & --3.762 & 1.531 & 7.403 & 0.403 & 0.577 &
0.700 \\
     5  & 8.359535 & 0.0000 & 0.7334 & 0.0000 & --3.840 & 1.548 & 7.415 & 0.301 & 0.679 &
0.700 \\
     6  & 8.414205 & 0.0000 & 0.6955 & 0.0000 & --3.900 & 1.577 & 7.429 & 0.203 & 0.778 &
0.700 \\
     7  & 8.453314 & 0.0000 & 0.6573 & 0.0000 & --3.930 & 1.631 & 7.449 & 0.105 & 0.876 &
0.700 \\
     8  & 8.476474 & 0.0000 & 0.6265 & 0.0000 & --3.885 & 1.736 & 7.485 & 0.027 & 0.954 &
0.700 \\
     9  & 8.482654 & 0.0000 & 0.6155 & 0.0000 & --2.735 & 2.336 & 7.549 & 0.000 & 0.981 &
0.700 \\
     10 & 8.483024 & 0.0000 & 0.6150 & 0.0000 & --1.793 & 2.684 & 7.518 & 0.000 & 0.981 &
0.700 \\
     11 & 8.483849 & 0.0000 & 0.6147 & 0.0000 & --1.099 & 2.981 & 7.530 & 0.000 & 0.981 &
0.700 \\
     12 & 8.484616 & 0.0000 & 0.6145 & 0.0000 & --0.663 & 3.214 & 7.573 & 0.000 & 0.981 &
0.700 \\
     13 & 8.485094 & 0.0000 & 0.6144 & 0.0000 & --0.403 & 3.377 & 7.616 & 0.000 & 0.981 &
0.700 \\
     14 & 8.485591 & 0.0047 & 0.6142 & 0.0000 & --0.123 & 3.570 & 7.678 & 0.000 & 0.981 &
0.700 \\
     15 & 8.485871 & 0.1401 & 0.6142 & 0.0000 &   0.042 & 3.690 & 7.719 & 0.000 & 0.981 &
0.700 \\
     16 & 8.485988 & 0.3395 & 0.6139 & 0.0000 &   0.114 & 3.742 & 7.737 & 0.000 & 0.981 &
0.700 \\
     17 & 8.486360 & 1.2953 & 0.6134 & 0.0000 &   0.347 & 3.902 & 7.792 & 0.000 & 0.981 &
0.700 \\
     18 & 8.486807 & 1.9842 & 0.6126 & 0.0000 &   0.636 & 4.084 & 7.851 & 0.000 & 0.981 &
0.699 \\
     19 & 8.487289 & 2.3994 & 0.6086 & 0.0000 &   0.952 & 4.272 & 7.912 & 0.000 & 0.981 &
0.690 \\
     20 & 8.487777 & 2.5843 & 0.5994 & 0.0000 &   1.276 & 4.455 & 7.973 & 0.000 & 0.981 &
0.674 \\
     21 & 8.488275 & 2.6446 & 0.5488 & 0.0912 &   1.035 & 4.535 & 8.068 & 0.000 & 0.979 &
0.666 \\
     22 & 8.504502 & 0.8827 & 0.5559 & 0.2191 & --0.126 & 4.196 & 8.088 & 0.000 & 0.833 &
0.666 \\
     23 & 8.565201 & 0.5946 & 0.6420 & 0.3891 & --0.771 & 4.056 & 8.152 & 0.000 & 0.255 &
0.666 \\
     24 & 8.584819 & 0.9931 & 0.6675 & 0.5042 & --1.099 & 3.948 & 8.164 & 0.000 & 0.134 &
0.666 \\
     25 & 8.584976 & 0.7203 & 0.6677 & 0.5146 & --0.958 & 4.019 & 8.174 & 0.000 & 0.128 &
0.666 \\
     26 & 8.589320 & 1.4878 & 0.6730 & 0.5088 & --0.668 & 4.269 & 8.264 & 0.000 & 0.011 &
0.666 \\
     27 & 8.589956 & 2.0177 & 0.6743 & 0.5076 & --0.282 & 4.511 & 8.327 & 0.000 & 0.000 &
0.666 \\
     28 & 8.590088 & 2.2555 & 0.6748 & 0.5068 &   0.525 & 4.800 & 8.336 & 0.000 & 0.000 &
0.666 \\
     29 & 8.590297 & 2.1974 & 0.6753 & 0.5054 &   1.236 & 4.966 & 8.308 & 0.000 & 0.000 &
0.666 \\
     30 & 8.591041 & 2.3732 & 0.6759 & 0.5051 &   2.563 & 5.300 & 8.320 & 0.000 & 0.000 &
0.666 \\
     31 & 8.591345 & 2.4422 & 0.6760 & 0.5079 &   3.156 & 5.455 & 8.345 & 0.000 & 0.000 &
0.666 \\
     32 & 8.591604 & 2.4787 & 0.6760 & 0.5278 &   4.022 & 5.649 & 8.375 & 0.000 & 0.000 &
0.666 \\
     33 & 8.591773 & 2.4978 & 0.6761 & 0.5467 &   4.954 & 5.773 & 8.370 & 0.000 & 0.000 &
0.666 \\
     34 & 8.591951 & 2.5098 & 0.6761 & 0.5718 &   6.546 & 5.922 & 8.347 & 0.000 & 0.000 &
0.666 \\
     35 & 8.592089 & 2.5156 & 0.6761 & 0.5993 &   8.768 & 6.066 & 8.311 & 0.000 & 0.000 &
0.666 \\
     36 & 8.592201 & 2.5179 & 0.6761 & 0.6275 &  11.911 & 6.204 & 8.261 & 0.000 & 0.000 &
0.666 \\
     37 & 8.592303 & 2.5184 & 0.6767 & 0.6538 &  16.623 & 6.347 & 8.200 & 0.000 & 0.000 &
0.666 \\
     38 & 8.592452 & 2.5109 & 0.6874 & 0.6769 &  26.118 & 6.519 & 8.100 & 0.000 & 0.000 &
0.666 \\
     39 & 8.592653 & 2.4887 & 0.7105 & 0.7044 &  38.345 & 6.663 & 8.012 & 0.000 & 0.000 &
0.666 \\
     40 & 8.592989 & 2.4352 & 0.7645 & 0.7611 &  54.315 & 6.844 & 7.956 & 0.000 & 0.000 &
0.666 \\
     41 & 8.593295 & 2.3695 & 0.8304 & 0.8284 &  67.708 & 7.021 & 7.950 & 0.000 & 0.000 &
0.666 \\
     42 & 8.593473 & 2.3217 & 0.8783 & 0.8769 &  76.249 & 7.145 & 7.960 & 0.000 & 0.000 &
0.666 \\
     43 & 8.593629 & 2.2723 & 0.9277 & 0.9267 &  84.732 & 7.276 & 7.977 & 0.000 & 0.000 &
0.666 \\
     44 & 8.593802 & 2.2072 & 0.9928 & 0.9922 &  96.061 & 7.454 & 8.006 & 0.000 & 0.000 &
0.666 \\
     45 & 8.593951 & 2.1427 & 1.0573 & 1.0569 & 108.209 & 7.645 & 8.040 & 0.000 & 0.000 &
0.666 \\
     46 & 8.594140 & 2.0504 & 1.1496 & 1.1494 & 128.357 & 7.955 & 8.100 & 0.000 & 0.000 &
0.666 \\
     47 & 8.594336 & 1.9444 & 1.2556 & 1.2555 & 158.486 & 8.419 & 8.198 & 0.000 & 0.000 &
0.666 \\
     48 & 8.594534 & 1.8282 & 1.3718 & 1.3718 & 203.220 & 9.363 & 8.448 & 0.000 & 0.000 &
0.666 \\
     \enddata
     
     \tablecomments{Table \ref{intmod} is published in its entirety in the electronic edition of the {\it Astrophysical Journal Supplement}. A portion is shown here for guidance regarding its form and content.}
     
\end{deluxetable}

\newcounter{tbl1}

\begin{deluxetable}{rrrrrrrrrr}    
     \tabletypesize{\footnotesize}
     \tablewidth{0pt}
     \tablecolumns{10}
     \tablecaption{Global properties of initial models\label{glbmod}}
     
     \tablehead{
          \colhead{$k$} & \colhead{$\log R$} & \colhead{$\log T_{\rm e}$} & \colhead{$\log
L$} &
          \colhead{$\log L_{\rm H}$} & \colhead{$\log L_{\rm He}$} & \colhead{$\log L_Z$} &
\colhead{$\log
               |L_\nu|$} & \colhead{$\log |L_{\rm th}|$} & \colhead{$I/(MR^2)$} \\
          \colhead{} & \colhead{$R_\sun$} & \colhead{K} & \colhead{$L_\sun$} &
\colhead{$L_\sun$} &
          \colhead{$L_\sun$} & \colhead{$L_\sun$} & \colhead{$L_\sun$} & \colhead{$L_\sun$}
& \colhead{}
     }
     \startdata
     \cutinhead{\normalsize $3.2000\ M_\sun$}
     1  & 0.3185 & 4.1067 & 2.0169 & 2.046 & --24.817 & \nodata\phn & 0.850* & --0.742* &
0.0532 \\*
     2  & 0.3700 & 4.0943 & 2.0706 & 2.100 & --24.468 & \nodata\phn & 0.907* & --0.856* &
0.0484 \\
     3  & 0.4211 & 4.0816 & 2.1220 & 2.151 & --24.069 & \nodata\phn & 0.960* & --0.793* &
0.0441 \\
     4  & 0.4748 & 4.0663 & 2.1680 & 2.197 & --23.649 & \nodata\phn & 1.004* & --1.622  &
0.0401 \\
     5  & 0.5367 & 4.0458 & 2.2100 & 2.239 & --23.167 & \nodata\phn & 1.045* & --1.508  &
0.0363 \\
     6  & 0.6027 & 4.0210 & 2.2426 & 2.271 & --22.622 & \nodata\phn & 1.076* & --1.371  &
0.0330 \\
     7  & 0.6743 & 3.9915 & 2.2677 & 2.296 & --21.864 & \nodata\phn & 1.100* & --0.936  &
0.0301 \\
     8  & 0.7196 & 3.9756 & 2.2946 & 2.322 & --20.516 & \nodata\phn & 1.126* & --0.341  &
0.0281 \\
     9  & 0.6532 & 4.0319 & 2.3872 & 2.457 & --17.619 & \nodata\phn & 1.261* &   1.382* &
0.0272 \\
     10 & 0.7516 & 3.9851 & 2.3971 & 2.459 & --18.162 & \nodata\phn & 1.264* &   1.303* &
0.0245 \\
     11 & 0.8540 & 3.9354 & 2.4027 & 2.452 & --17.556 & \nodata\phn & 1.256* &   1.087* &
0.0222 \\
     12 & 0.9536 & 3.8789 & 2.3763 & 2.436 & --16.110 & \nodata\phn & 1.240* &   1.241* &
0.0209 \\
     13 & 1.0493 & 3.8198 & 2.3313 & 2.410 & --14.687 & \nodata\phn & 1.214* &   1.416* &
0.0203 \\
     14 & 1.1528 & 3.7388 & 2.2140 & 2.346 & --12.753 & \nodata\phn & 1.151* &   1.644* &
0.0277 \\
     15 & 1.1434 & 3.7164 & 2.1058 & 2.282 & --11.493 & \nodata\phn & 1.087* &   1.714* &
0.0589 \\
     16 & 1.1516 & 3.7078 & 2.0877 & 2.256 & --10.951 & \nodata\phn & 1.061* &   1.668* &
0.0786 \\
     17 & 1.2552 & 3.6859 & 2.2070 & 2.262 &  --9.345 & \nodata\phn & 1.067* &   1.008* &
0.1163 \\
     18 & 1.3566 & 3.6719 & 2.3542 & 2.376 &  --7.553 & \nodata\phn & 1.181* &   0.538  &
0.1266 \\
     19 & 1.4598 & 3.6592 & 2.5096 & 2.524 &  --5.262 & \nodata\phn & 1.329* &   1.017  &
0.1313 \\
     20 & 1.5650 & 3.6465 & 2.6692 & 2.682 &  --2.223 & \nodata\phn & 1.486* &   1.232  &
0.1348 \\
     21 & 1.6659 & 3.6338 & 2.8203 & 2.831 &    1.541 & --35.172    & 1.636* &   0.907* &
0.1353 \\
     22 & 1.2213 & 3.6949 & 2.1755 & 2.096 &    1.517 & --30.489    & 0.907* & --0.844  &
0.1067 \\
     23 & 1.2919 & 3.6923 & 2.3061 & 2.064 &    1.975 & --26.020    & 0.885* & --0.692* &
0.0836 \\
     24 & 1.3626 & 3.6804 & 2.4001 & 1.978 &    2.279 & --25.925    & 0.809* &   1.437* &
0.0956 \\
     25 & 1.3140 & 3.6883 & 2.3344 & 1.974 &    2.065 & --25.309    & 0.808* &   1.085  &
0.0882 \\
     26 & 1.3971 & 3.6729 & 2.4389 & 2.100 &    2.188 & --20.796    & 0.990* &   0.644  &
0.1107 \\
     27 & 1.4839 & 3.6604 & 2.5628 & 2.354 &    1.841 & --17.700    & 1.276* &   1.950  &
0.1186 \\
     28 & 1.5682 & 3.6492 & 2.6864 & 2.452 &    2.221 & --16.814    & 1.393* &   1.786  &
0.1202 \\
     29 & 1.5422 & 3.6526 & 2.6483 & 2.142 &    2.477 & --17.804    & 1.166* &   1.320  &
0.1199 \\
     30 & 1.6411 & 3.6395 & 2.7933 & 1.145 &    2.767 & --16.494    & 1.057* &   1.537  &
0.1211 \\
     31 & 1.7359 & 3.6268 & 2.9323 & 0.784 &    2.896 & --14.984    & 1.255* &   1.909  &
0.1208 \\
     32 & 1.8392 & 3.6129 & 3.0832 & 1.104 &    3.061 & --12.963    & 1.582* &   1.938  &
0.1197 \\
     33 & 1.9346 & 3.5998 & 3.2218 & 0.643 &    3.203 & --12.489    & 1.706* &   2.065  &
0.1185 \\
     34 & 2.0384 & 3.5855 & 3.3722 & 0.503 &    3.350 & --12.059    & 1.874* &   2.271  &
0.1172 \\
     35 & 2.1341 & 3.5722 & 3.5104 & 1.300 &    3.486 & --11.376    & 2.063* &   2.433  &
0.1164 \\
     36 & 2.2172 & 3.5607 & 3.6303 & 2.633 &    3.699 & --10.606    & 2.294* &   2.983* &
0.1164 \\
     37 & 2.3138 & 3.5472 & 3.7698 & 3.420 &    3.490 &  --9.932    & 2.599* &   2.748  &
0.1176 \\
     38 & 2.4199 & 3.5325 & 3.9230 & 3.840 &    3.224 & --10.439    & 2.816* &   2.634  &
0.1211 \\
     39 & 2.5175 & 3.5186 & 4.0627 & 4.006 &    3.526 & --11.899    & 2.892* &   3.067* &
0.1268 \\
     40 & 2.6174 & 3.5037 & 4.2028 & 4.152 &    3.371 & --12.642    & 2.998* &   2.606  &
0.1354 \\
     41 & 2.7032 & 3.4899 & 4.3192 & 4.269 &    3.476 & --11.897    & 3.108* &   2.756  &
0.1457 \\
     42 & 2.7587 & 3.4807 & 4.3932 & 4.342 &    3.548 & --11.149    & 3.181* &   2.852  &
0.1543 \\
     43 & 2.8134 & 3.4716 & 4.4663 & 4.415 &    3.637 & --10.325    & 3.255* &   2.859  &
0.1643 \\
     44 & 2.8693 & 3.4629 & 4.5435 & 4.491 &    3.727 &  --9.325    & 3.333* &   2.914  &
0.1752 \\
     45 & 2.9066 & 3.4580 & 4.5984 & 4.544 &    3.739 &  --8.532    & 3.389* &   3.221  &
0.1820 \\
     46 & 2.9421 & 3.4546 & 4.6560 & 4.597 &    3.817 &  --7.579    & 3.445* &   3.297  &
0.1869 \\
     47 & 2.9703 & 3.4535 & 4.7080 & 4.640 &    3.829 &  --6.436    & 3.496* &   3.578  &
0.1887 \\
     48 & 2.9993 & 3.4538 & 4.7669 & 4.665 &    3.838 &    1.350    & 3.566* &   3.956  &
0.1899 \\
     \enddata
     
     \tablecomments{Table \ref{glbmod} is published in its entirety in the electronic edition of the {\it Astrophysical Journal Supplement}. A portion is shown here for guidance regarding its form and content.}
     
\end{deluxetable}

\newcounter{tbl2}

Table~\ref{intmod} is arranged in segments, by stellar mass, $M_{\rm i}$. Successive columns list:
\begin{list}{(\arabic{tbl1})}{\usecounter{tbl1}}
     \item $k$ --- mass loss sequence number;
     \item $\log t$ --- age (yr);
     \item $M_{\rm ce}$ --- mass of the convective envelope ($M_\sun$);
     \item $M_{\rm c}$ --- core mass ($M_\sun$);
     \item $M_{\rm ic}$ --- inner core mass ($M_\sun$);
     \item $\psi_{\rm c}$ --- central electron chemical potential ($\mu_e$, in units of $kT$);
     \item $\log \rho_{\rm c}$ --- central density (${\rm g\,cm^{-3}}$);
     \item $\log T_{\rm c}$ --- central temperature (K);
     \item $X_{\rm c}$ --- central hydrogen abundance (fraction by mass);
     \item $Y_{\rm c}$ --- central helium abundance (fraction by mass); and
     \item $X_{\rm s}$ --- surface hydrogen abundance (fraction by mass).
\end{list}

Age $\log t$ is measured from the ZAMS model (excluding pre-main-sequence evolution). The mass of the convective envelope $M_{\rm ce}$ refers to the mass depth of the base of the outermost convection zone. The core mass $M_{\rm c}$ refers to the mass coordinate at which the helium abundance is halfway between the surface helium abundance and the maximum helium abundance in the stellar interior. The inner core mass $M_{\rm ic}$ identifies the mass coordinate at which the helium abundance is halfway between the maximum helium abundance in the stellar interior and the minimum helium abundance interior to that maximum; in the absence of measurable helium depletion in the hydrogen-exhausted core, $M_{\rm ic}$ is set to a default value of zero. $M_{\rm c}$ and $M_{\rm ic}$ characterize the \emph{range} in mass over which hydrogen and helium are being depleted during their respective core burning phases, and \emph{not} the amount of mass that has been consumed. Upon core fuel exhaustion, $M_{\rm c}$ and $M_{\rm ic}$ mark the midpoints in hydrogen and helium depletion profiles, respectively. The dimensionless central electron chemical potential $\psi_{\rm c}$ measures the degree of electron degeneracy ($\psi_{\rm c} > 0$).

Like Table~\ref{intmod}, Table~\ref{glbmod} is arranged in segments, by stellar mass, $M_{\rm i}$. Successive columns list:
\begin{list}{(\arabic{tbl2})}{\usecounter{tbl2}}
     \item $k$ --- mass loss sequence number;
     \item $\log R$ --- radius ($R_\sun$);
     \item $\log T_{\rm e}$ --- effective temperature (K);
     \item $\log L$ --- stellar luminosity ($L_\sun$);
     \item $\log L_{\rm H}$ --- hydrogen-burning luminosity ($L_\sun$);
     \item $\log L_{\rm He}$ --- helium-burning luminosity ($L_\sun$);
     \item $\log L_Z$ --- heavy-element (carbon-, oxygen-, etc.) burning luminosity ($L_\sun$);
     \item $\log |L_{\nu}|$ --- log neutrino luminosity ($L_\sun$, with asterisk, *, appended to signify that this is a \emph{negative} contribution to the net stellar luminosity);
     \item $\log |L_{\rm th}|$ --- gravothermal luminosity ($L_\sun$, with asterisk, *, appended where the gravothermal luminosity is negative); and
     \item $I/(MR^2)$ --- dimensionless moment of inertia.
\end{list}

Table~\ref{dynml} summarizes the quantitative results of our investigation for both those model sequences derived from initial models with standard mixing-length convective envelopes (columns 2-7) and those sequences derived from initial models with artificially isentropic convective envelopes (columns 8-13). For each set of sequences, it identifies critical points marking the onset of runaway (dynamical timescale) mass transfer, and the (critical) initial conditions (mass-radius exponent and mass ratio) corresponding to those critical points. As noted in Section~\ref{sec_supad}, the onset of dynamical timescale mass transfer is not instantaneous but is preceded by an episode of accelerating mass transfer. We associate the transition to dynamical timescale mass transfer with the mass transfer rate equaling the nominal thermal timescale rate of the initial model of the sequence, $\dot{M}_{\rm KH} = - R_{\rm i} L_{\rm i}/GM_{\rm i}$; beyond that rate, the response of the donor becomes asymptotically adiabatic. We then define the critical mass ratio for dynamical mass transfer as the minimum initial mass ratio for which $\dot{M}$ reaches $\dot{M}_{\rm KH}$.

\begin{deluxetable}{rrrrrrrrrrrrrr}
     \tabletypesize{\footnotesize}
     %\rotate
     \tablewidth{0pt}
     \tablecolumns{14}
     \tablecaption{Thresholds for conservative dynamical timescale mass transfer\label{dynml}}
     
     \tablehead{
          \colhead{} & \multicolumn{6}{c}{Mixing-length convection} & \colhead{} &
          \multicolumn{6}{c}{Isentropic convection} \\[8pt]
          \cline{2-7} \cline{9-14}
          \colhead{$k$} & \colhead{$\log R_{\rm i}$} & \colhead{$M_{\rm KH}$} & \colhead{$\log R_{\rm KH}$} & \colhead{$\log R^*_{\rm KH}$} & \colhead{$\zeta_{\rm ad}$}
& \colhead{$q_{\rm ad}$} &
          & \colhead{$\Delta_{\rm exp}$} & \colhead{$\tilde{M}_{\rm KH}$} & \colhead{$\log
               \tilde{R}_{\rm KH}$} & \colhead{$\log \tilde{R}^*_{\rm KH}$} &
\colhead{$\tilde{\zeta}_{\rm ad}$}
          & \colhead{$\tilde{q}_{\rm ad}$} \\
          \colhead{} & \colhead{$R_\sun$} & \colhead{$M_\sun$} & \colhead{$R_\sun$} &
\colhead{$R_\sun$}
          & \colhead{} & \colhead{} & \colhead{} & \colhead{} & \colhead{$M_\sun$} &
\colhead{$R_\sun$} &
          \colhead{$R_\sun$} & \colhead{} & \colhead{}
     }
     \startdata
     \cutinhead{\normalsize $3.2000\ M_\sun$}
     1  & 0.3185 & 1.8842 & 0.0306 & 0.0315 & 3.873 & 2.592 & & 0.0002 & 1.8838 & 0.0306 &
0.0315 & 3.876 & 2.593 \\*
     2  & 0.3701 & 1.9145 & 0.0516 & 0.0527 & 4.227 & 2.757 & & 0.0002 & 1.9130 & 0.0516 &
0.0527 & 4.230 & 2.759 \\
     3  & 0.4211 & 1.9433 & 0.0730 & 0.0742 & 4.593 & 2.929 & & 0.0003 & 1.9423 & 0.0730 &
0.0742 & 4.597 & 2.931 \\
     4  & 0.4749 & 1.9722 & 0.0978 & 0.0991 & 4.977 & 3.109 & & 0.0002 & 1.9722 & 0.0978 &
0.0991 & 4.980 & 3.111 \\
     5  & 0.5368 & 2.0058 & 0.1295 & 0.1310 & 5.404 & 3.310 & & 0.0003 & 2.0056 & 0.1295 &
0.1310 & 5.407 & 3.311 \\
     6  & 0.6028 & 2.0411 & 0.1675 & 0.1692 & 5.836 & 3.513 & & 0.0002 & 2.0406 & 0.1674 &
0.1692 & 5.839 & 3.514 \\
     7  & 0.6745 & 2.0786 & 0.2129 & 0.2148 & 6.280 & 3.721 & & 0.0004 & 2.0781 & 0.2128 &
0.2148 & 6.285 & 3.724 \\
     8  & 0.7197 & 2.1043 & 0.2394 & 0.2416 & 6.615 & 3.879 & & 0.0005 & 2.1035 & 0.2393 &
0.2415 & 6.623 & 3.883 \\
     9  & 0.6533 & 2.0820 & 0.1647 & 0.1666 & 6.698 & 3.918 & & 0.0004 & 2.0814 & 0.1647 &
0.1666 & 6.704 & 3.921 \\
     10 & 0.7518 & 2.1333 & 0.2347 & 0.2370 & 7.274 & 4.190 & & 0.0004 & 2.1327 & 0.2346 &
0.2369 & 7.280 & 4.192 \\
     11 & 0.8540 & 2.1833 & 0.3134 & 0.3162 & 7.827 & 4.451 & & 0.0011 & 2.1821 & 0.3131 &
0.3160 & 7.846 & 4.459 \\
     12 & 0.9536 & 2.2380 & 0.4076 & 0.4111 & 8.134 & 4.595 & & 0.0029 & 2.2343 & 0.4067 &
0.4102 & 8.188 & 4.621 \\
     13 & 1.0493 & 2.2827 & 0.5098 & 0.5141 & 8.217 & 4.634 & & 0.0056 & 2.2763 & 0.5082 &
0.5124 & 8.321 & 4.684 \\
     14 & 1.1528 & 2.3407 & 0.7061 & 0.7119 & 6.804 & 3.968 & & 0.0077 & 2.3318 & 0.7039 &
0.7097 & 6.941 & 4.032 \\
     15 & 1.1434 & 3.1157 & 1.1164 & 1.1257 & 2.488 & 1.944 & & 0.0051 & 3.0876 & 1.1072 &
1.1166 & 2.898 & 2.136 \\
     16 & 1.1516 & 3.0639 & 1.1322 & 1.1409 & 1.165 & 1.329 & & 0.0048 & 3.0072 & 1.1247 &
1.1334 & 1.394 & 1.435 \\
     17 & 1.2552 & 2.9827 & 1.2476 & 1.2581 & 0.381 & 0.965 & & 0.0057 & 2.8870 & 1.2458 &
1.2566 & 0.550 & 1.043 \\
     18 & 1.3566 & 2.9428 & 1.3509 & 1.3647 & 0.306 & 0.930 & & 0.0073 & 2.8285 & 1.3505 &
1.3648 & 0.493 & 1.017 \\
     19 & 1.4598 & 2.9004 & 1.4545 & 1.4728 & 0.298 & 0.926 & & 0.0096 & 2.7698 & 1.4549 &
1.4740 & 0.514 & 1.026 \\
     20 & 1.5650 & 2.8547 & 1.5598 & 1.5844 & 0.305 & 0.930 & & 0.0128 & 2.7078 & 1.5613 &
1.5872 & 0.557 & 1.046 \\
     21 & 1.6659 & 2.8037 & 1.6596 & 1.6926 & 0.343 & 0.947 & & 0.0169 & 2.6376 & 1.6620 &
1.6969 & 0.641 & 1.085 \\
     22 & 1.2213 & 3.0030 & 1.2112 & 1.2209 & 0.502 & 1.021 & & 0.0053 & 2.9159 & 1.2079 &
1.2179 & 0.676 & 1.101 \\
     23 & 1.2919 & 3.0116 & 1.2720 & 1.2847 & 0.920 & 1.215 & & 0.0069 & 2.9325 & 1.2648 &
1.2776 & 1.161 & 1.327 \\
     24 & 1.3626 & 2.9703 & 1.3466 & 1.3615 & 0.667 & 1.098 & & 0.0080 & 2.8712 & 1.3413 &
1.3567 & 0.899 & 1.205 \\
     25 & 1.3140 & 2.9970 & 1.2958 & 1.3091 & 0.806 & 1.162 & & 0.0072 & 2.9105 & 1.2893 &
1.3028 & 1.040 & 1.271 \\
     26 & 1.3971 & 2.9415 & 1.3861 & 1.4021 & 0.468 & 1.005 & & 0.0085 & 2.8283 & 1.3835 &
1.3999 & 0.686 & 1.106 \\
     27 & 1.4839 & 2.9008 & 1.4743 & 1.4944 & 0.413 & 0.980 & & 0.0106 & 2.7729 & 1.4730 &
1.4939 & 0.652 & 1.090 \\
     28 & 1.5682 & 2.8524 & 1.5565 & 1.5831 & 0.455 & 0.999 & & 0.0133 & 2.7161 & 1.5556 &
1.5833 & 0.721 & 1.123 \\
     29 & 1.5422 & 2.8658 & 1.5312 & 1.5557 & 0.444 & 0.994 & & 0.0124 & 2.7327 & 1.5301 &
1.5557 & 0.699 & 1.112 \\
     30 & 1.6411 & 2.8204 & 1.6282 & 1.6607 & 0.482 & 1.012 & & 0.0162 & 2.6714 & 1.6275 &
1.6616 & 0.784 & 1.152 \\
     31 & 1.7359 & 2.7727 & 1.7198 & 1.7626 & 0.546 & 1.041 & & 0.0212 & 2.6091 & 1.7193 &
1.7645 & 0.905 & 1.208 \\
     32 & 1.8392 & 2.7169 & 1.8181 & 1.8764 & 0.638 & 1.084 & & 0.0286 & 2.5371 & 1.8177 &
1.8798 & 1.078 & 1.288 \\
     33 & 1.9346 & 2.6648 & 1.9074 & 1.9848 & 0.740 & 1.131 & & 0.0381 & 2.4688 & 1.9072 &
1.9906 & 1.280 & 1.382 \\
     34 & 2.0384 & 2.6037 & 2.0024 & 2.1085 & 0.876 & 1.195 & & 0.0527 & 2.3909 & 2.0026 &
2.1185 & 1.565 & 1.515 \\
     35 & 2.1341 & 2.5415 & 2.0866 & 2.2302 & 1.041 & 1.271 & & 0.0722 & 2.3143 & 2.0876 &
2.2473 & 1.924 & 1.682 \\
     36 & 2.2172 & 2.4827 & 2.1545 & 2.3452 & 1.243 & 1.365 & & 0.0972 & 2.2472 & 2.1569 &
2.3736 & 2.375 & 1.892 \\
     37 & 2.3138 & 2.4086 & 2.2197 & 2.4962 & 1.641 & 1.550 & & 0.1692 & 2.1674 & 2.2400 &
2.5846 & 3.386 & 2.364 \\
     38 & 2.4199 & 2.3565 & 2.2531 & 2.6922 & 2.558 & 1.977 & & 0.2347 & 2.1407 & 2.2486 &
2.8188 & 5.565 & 3.385 \\
     39 & 2.5175 & 2.3739 & 2.1929 & 2.9100 & 4.883 & 3.065 & & 0.2135 & 2.2079 & 2.1435 &
2.9644 & 8.764 & 4.893 \\
     40 & 2.6173 & 2.4569 & 1.5274 & 3.2061 &25.63  &12.93  & & 0.2991 & 2.4039 & 1.4208 &
3.2451 &43.43  &21.51 \\
     41 & 2.7032 &\nodata\phn &\nodata\phn &\nodata\phn &\nodata\phn &\nodata\phn &
&\nodata\phn &\nodata\phn &\nodata\phn &\nodata\phn &\nodata\phn &\nodata\phn \\
     42 & 2.7587 &\nodata\phn &\nodata\phn &\nodata\phn &\nodata\phn &\nodata\phn &
&\nodata\phn &\nodata\phn &\nodata\phn &\nodata\phn &\nodata\phn &\nodata\phn \\
     43 & 2.8134 &\nodata\phn &\nodata\phn &\nodata\phn &\nodata\phn &\nodata\phn &
&\nodata\phn &\nodata\phn &\nodata\phn &\nodata\phn &\nodata\phn &\nodata\phn \\
     44 & 2.8693 &\nodata\phn &\nodata\phn &\nodata\phn &\nodata\phn &\nodata\phn &
&\nodata\phn &\nodata\phn &\nodata\phn &\nodata\phn &\nodata\phn &\nodata\phn \\
     45 & 2.9066 &\nodata\phn &\nodata\phn &\nodata\phn &\nodata\phn &\nodata\phn &
&\nodata\phn &\nodata\phn &\nodata\phn &\nodata\phn &\nodata\phn &\nodata\phn \\
     46 & 2.9422 &\nodata\phn &\nodata\phn &\nodata\phn &\nodata\phn &\nodata\phn &
&\nodata\phn &\nodata\phn &\nodata\phn &\nodata\phn &\nodata\phn &\nodata\phn \\
     47 & 2.9704 &\nodata\phn &\nodata\phn &\nodata\phn &\nodata\phn &\nodata\phn &
&\nodata\phn &\nodata\phn &\nodata\phn &\nodata\phn &\nodata\phn &\nodata\phn \\
     48 & 2.9993 &\nodata\phn &\nodata\phn &\nodata\phn &\nodata\phn &\nodata\phn &
&\nodata\phn &\nodata\phn &\nodata\phn &\nodata\phn &\nodata\phn &\nodata\phn \\
     \enddata
     
     \tablecomments{Table \ref{dynml} is published in its entirety in the electronic edition of the {\it Astrophysical Journal Supplement}. A portion is shown here for guidance regarding its form and content.}
     
\end{deluxetable}

Like Tables~\ref{intmod} and~\ref{glbmod}, Table~\ref{dynml} is arranged in segments, by stellar mass, $M_{\rm i}$. Successive columns list:
\newcounter{tbl3}
\begin{list}{(\arabic{tbl3})}{\usecounter{tbl3}}
     \item $k$ --- mass-loss sequence number;
\end{list}
for models with standard mixing-length convective envelopes:
\begin{list}{(\arabic{tbl3})}{\usecounter{tbl3}}
     \setcounter{tbl3}{1}
     \item $\log R_{\rm i}$ --- initial radius ($R_\sun$);
     \item $M_{\rm KH}$ --- mass threshold at which $\dot{M} = - M/\tau_{\rm KH}$;
     \item $\log R_{\rm KH}$ --- Roche lobe radius at which $\dot{M} = - M/\tau_{\rm KH}$;
     \item $\log R_{\rm KH}^*$ --- stellar radius when $\dot{M} = - M/\tau_{\rm KH}$;
     \item $\zeta_{\rm ad}$ --- critical mass-radius exponent for dynamical timescale mass transfer;
     \item $q_{\rm ad}$ --- critical mass ratio for dynamical timescale (conservative) mass transfer;
\end{list}
and for models with artificially isentropic convective envelopes:
\begin{list}{(\arabic{tbl3})}{\usecounter{tbl3}}
     \setcounter{tbl3}{7}
     \item $\Delta_{\rm exp} \equiv \log (\tilde{R}_i/R_i)$ --- superadiabatic expansion factor;
     \item $\tilde{M}_{\rm KH}$ --- mass threshold at which $\dot{M} = - M/\tau_{\rm KH}$;
     \item $\log \tilde{R}_{\rm KH}$ --- Roche lobe radius at which $\dot{M} = - M/\tau_{\rm KH}$;
     \item $\log \tilde{R}_{\rm KH}^*$ --- stellar radius when $\dot{M} = - M/\tau_{\rm KH}$;
     \item $\tilde{\zeta}_{\rm ad}$ --- critical mass-radius exponent for dynamical timescale mass transfer;
     and
     \item $\tilde{q}_{\rm ad}$ --- critical mass ratio for dynamical timescale (conservative) mass transfer.
\end{list}

Columns (3)-(5) and (9)-(11) refer to the points in the critical mass loss sequences at which $\dot{M}$ just reaches $\dot{M}_{\rm KH}$, the characteristic mass loss rate which we identify with the transition from thermal to dynamical timescale mass transfer. The corresponding initial conditions leading to these critical points are found in columns (6) and (7) ($\zeta_{\rm ad}$ and $q_{\rm ad}$, respectively) for the mixing length convection models, and in columns (12) and (13) ($\tilde{\zeta}_{\rm ad}$ and $\tilde{q}_{\rm ad}$, respectively) for the isentropic convection models. For reasons outlined above, we consider the critical mass-radius exponents and mass ratios for these models, $\tilde{\zeta}_{\rm ad}$ and $\tilde{q}_{\rm ad}$ respectively the more realistic, and adopt them in preference to $\zeta_{\rm ad}$ and $q_{\rm ad}$ below in applying our results to real systems.

\begin{figure}
\centering
\includegraphics[scale=0.36]{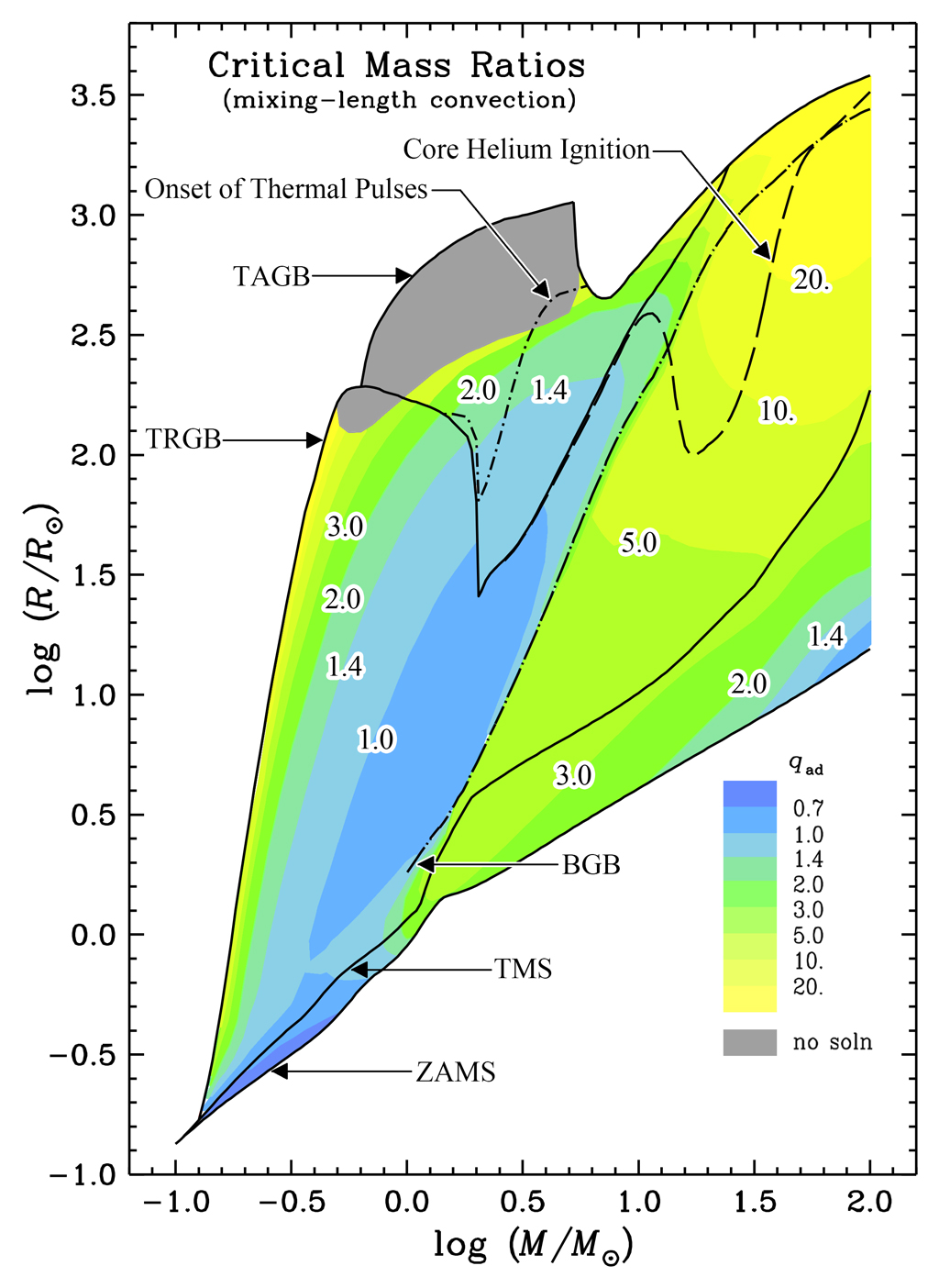}
\caption{Critical mass ratios, $q_{\rm ad}$, for the onset of dynamical timescale mass transfer as derived from standard evolutionary models, in the mass-radius diagram.\label{qcrit_real}}
\end{figure}

\begin{figure}
\centering
\includegraphics[scale=0.36]{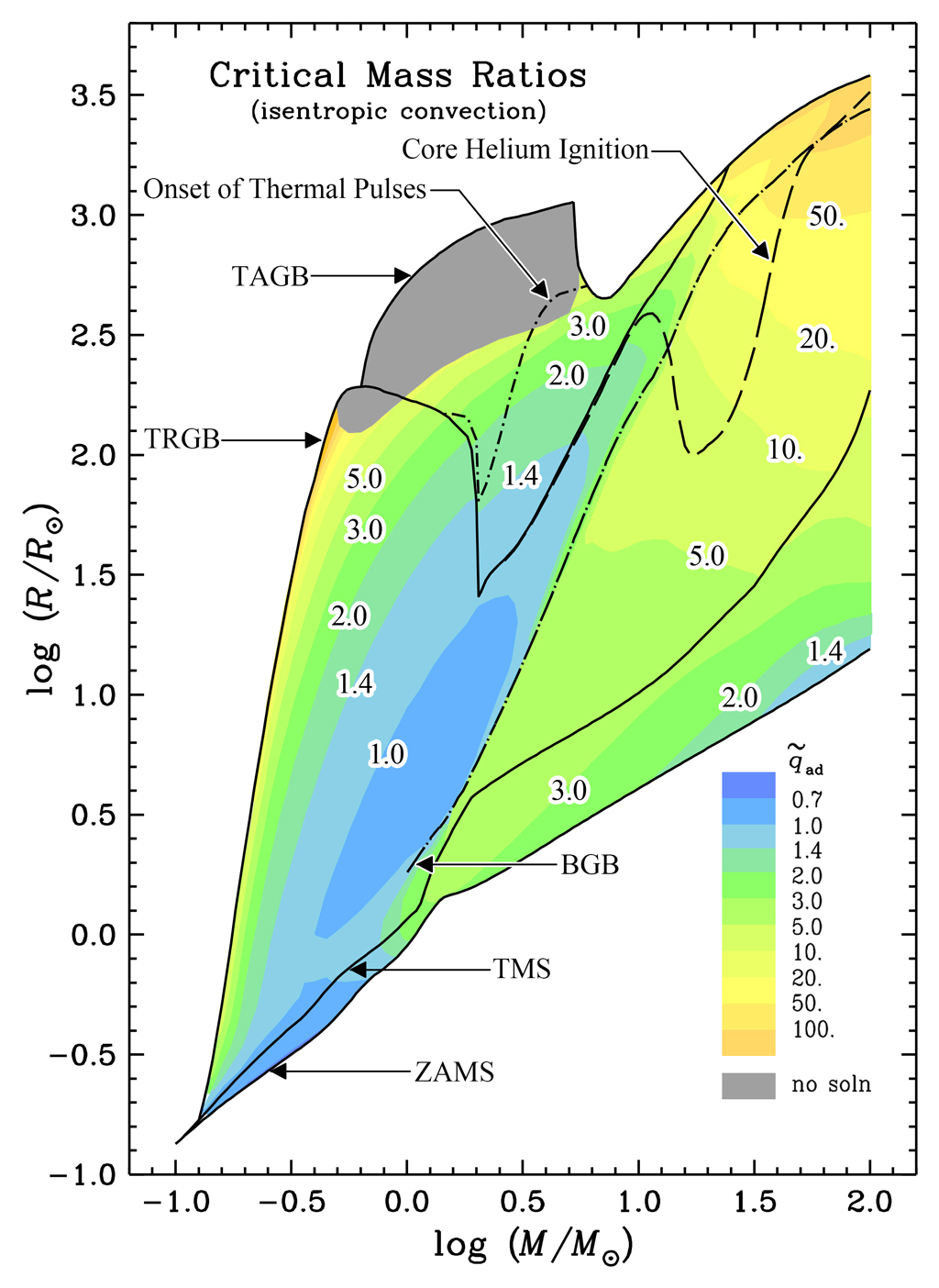}
\caption{Critical mass ratios, $\tilde{q}_{\rm ad}$, for the onset of dynamical timescale mass transfer as derived from modified evolutionary models with isentropic surface convection zones, in the mass-radius diagram.  These models mimic the effects of rapid thermal relaxation in the outer layers of convective stellar envelopes by suppressing the destabilizing effect superadiabatic expansion, thus providing more realistic estimates of critical mass ratios than the models shown in Figure~\ref{qcrit_real}.\label{qcrit_isen}}
\end{figure}

As we mentioned in Paper II, the slight difference in $\log R_{\rm i}$ between Table~\ref{dynml} for the mass-loss sequences and Table~\ref{glbmod} for the corresponding evolutionary models was caused by modifying the surface boundary conditions imposed on the adiabatic sequences (see Paper I). But we find that the difference in $\log R_{\rm i}$ is, in all cases, negligible in magnitude.

The critical mass ratios found in Table~\ref{dynml} are presented graphically in the form of contour plots in Figures~\ref{qcrit_real} and~\ref{qcrit_isen} for mixing-length and isentropic envelope models, respectively. For those MS and HG stars, similar to what we found in Paper II, the solutions for $q_{\rm ad}$ and $\tilde{q}_{\rm ad}$ differ very little from each other qualitatively, although $\tilde{q}_{\rm ad}$ is systematically larger than $q_{\rm ad}$. The difference is quantitatively small except for low-mass MS stars and massive stars. Figures~\ref{qcrit_real} and~\ref{qcrit_isen} display the nearly uniform trend toward larger critical mass ratios with larger radii in the MS and HG. These intermediate-mass and massive stars have very thin surface convection zones and typically contract very rapidly in response to adiabatic mass loss. Their critical mass ratios for dynamical timescale mass transfer are then set by the \emph{delayed dynamical instability}. 

For those RGB and AGB stars, the solutions for $q_{\rm ad}$ and $\tilde{q}_{\rm ad}$ also differ very little from each other qualitatively. $\tilde{q}_{\rm ad}$ is systematically larger than $q_{\rm ad}$, and the difference between them is also very small. Thermal relaxation of the donor star becomes increasingly important among luminous donors with extended envelopes. The convergence of the dynamical and thermal timescales of those thermal pulsation AGB stars is related to the growth of massive surface superadiabatic zones, and ultimately to the failure of adiabatic mass loss models (gray zones in Figures~\ref{qcrit_real} and~\ref{qcrit_isen}), as the underlying convective envelope becomes energetically unbound. We will present a short discussion in the last section.

\section{COMPARISON of \lowercase{$q_{\rm cr}$} from ADIABATIC MASS LOSS MODELS vs. PREVIOUS RESULTS}
\label{comparison}

We compare the dynamical mass transfer threshold mass-radius exponents $\zeta_{\rm ad}$ and/or corresponding limiting mass ratio $q_{\rm ad}$ derived from our calculations with those from polytropic models ($\zeta_{\rm ad}$), and time-dependent calculations ($q_{\rm cr}$) in this section.

\begin{figure}
\centering
\includegraphics[scale=0.4]{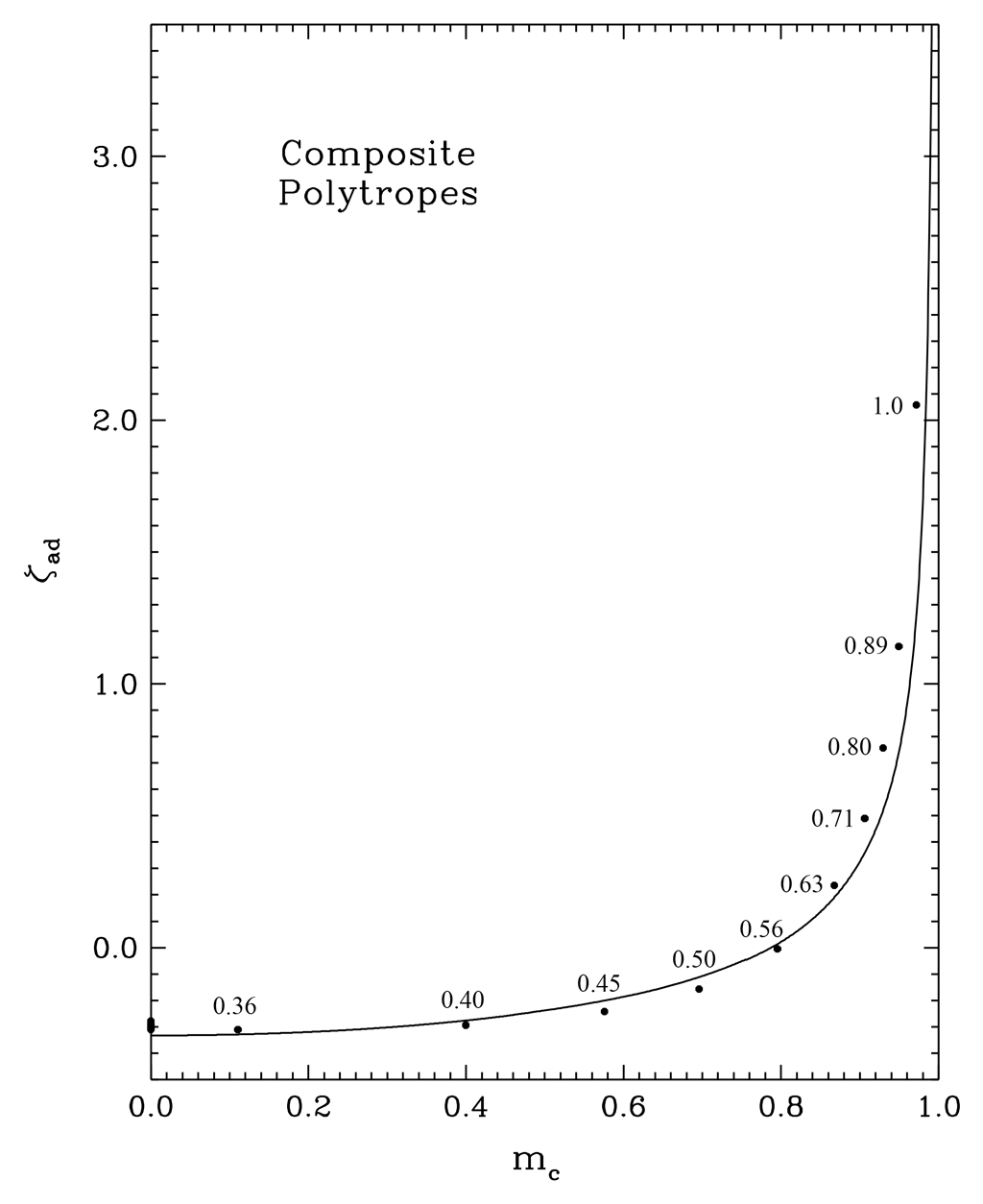}
\caption{Threshold mass-radius exponent, $\zeta_{\rm ad}$, for the onset of dynamical timescale mass transfer as derived from this paper (dots) and composite polytropes (solid line), as function of the core mass function ($m_{\rm c}\equiv (M-M_{\rm ce})/M$) of MS stars. The mass of the stellar models are labeled and they are correlated to the $\zeta_{\rm ad}$ derived from our ZAMS models. Composite polytropes give a good approximation for low mass stars and show consistent results with our calculation. \label{qcrit_composite}}
\end{figure}

\begin{figure}
\centering
\includegraphics[scale=0.6]{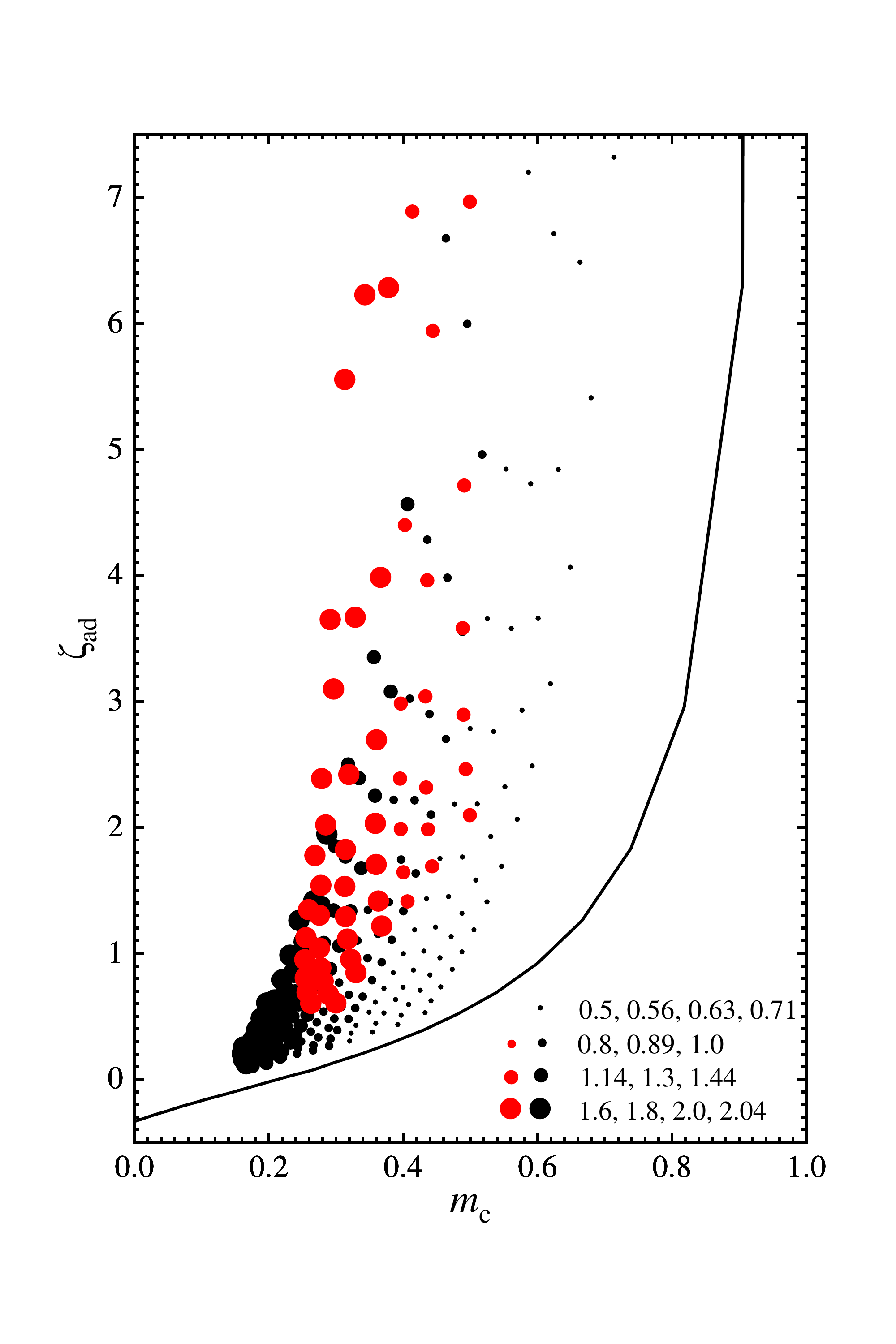}
\caption{Threshold mass-radius exponent, $\zeta_{\rm ad}$, for the onset of dynamical timescale mass transfer as derived from this paper (dots) and condensed polytropes (solid line), as a function of the core mass function ($m_{\rm c}$) of GB stars. Black dots show results from RGB stars, and red dots derived from AGB stars. Four groups of the initial mass, which equal to (0.50, 0.56, 0.63, and 0.71), (0.80, 0.89, and 1.00), (1.14, 1.30, and 1.44), and (1.60, 1.80, 2.00, and 2.04), are shown by dots of four different sizes from small to large, respectively. Condensed polytropes do not count into the non-idea gas effect and inefficient convection, and we see an extended pattern of $q_{\rm ad}$ from our calculation. \label{qcrit_condensed}}
\end{figure}

Useful insights into the behavior of donor stars under adiabatic mass loss have been studied by \citet{hjel87} from simplified polytropic stellar models. They investigated the properties of polytropic models with power-law equations of state, including complete polytropes, composite polytropes ($n = 3$ cores with $n = 3/2$ envelopes and $\gamma = 5/3$ equations of state), and condensed polytropes ($n = 3/2$ envelopes with point mass cores). Composite polytropic models are relevant for low mass ZAMS stars with mass $0.3 \le M/M_\odot \le 1 $. Condensed polytropic models are applicable to giant branch stars. As we expected, composite polytropes give a good approximation for low mass stars and show consistent results with our calculation in Figure~\ref{qcrit_composite}. However, it should be noted that condensed polytropic models are usually established under the assumptions that all gases involved are ideal gases, and that convection is efficient; consequently, in these models, $q_{\rm ad}$ tends to vary as a simple function of core mass function, $m_{\rm c}\equiv (M-M_{\rm ce})/M$. Our results, which are not derived under such assumptions, exhibit an extended pattern as a function of core mass function $m_{\rm c}$ for stars with different initial mass in Figure~\ref{qcrit_condensed}. Four groups of initial mass, which equals (0.50, 0.56, 0.63, and 0.71), (0.80, 0.89, and 1.00), (1.14, 1.30, and 1.44), and (1.60, 1.80, 2.00, and 2.04), are shown by dots of four different sizes from small to large, respectively. The donor stars on the RGB and AGB tend to be more dynamically stable (harder to suffer a dynamical timescale mass transfer) than we thought from polytropic models.

\begin{figure}
\centering
\includegraphics[angle=270,scale=0.6]{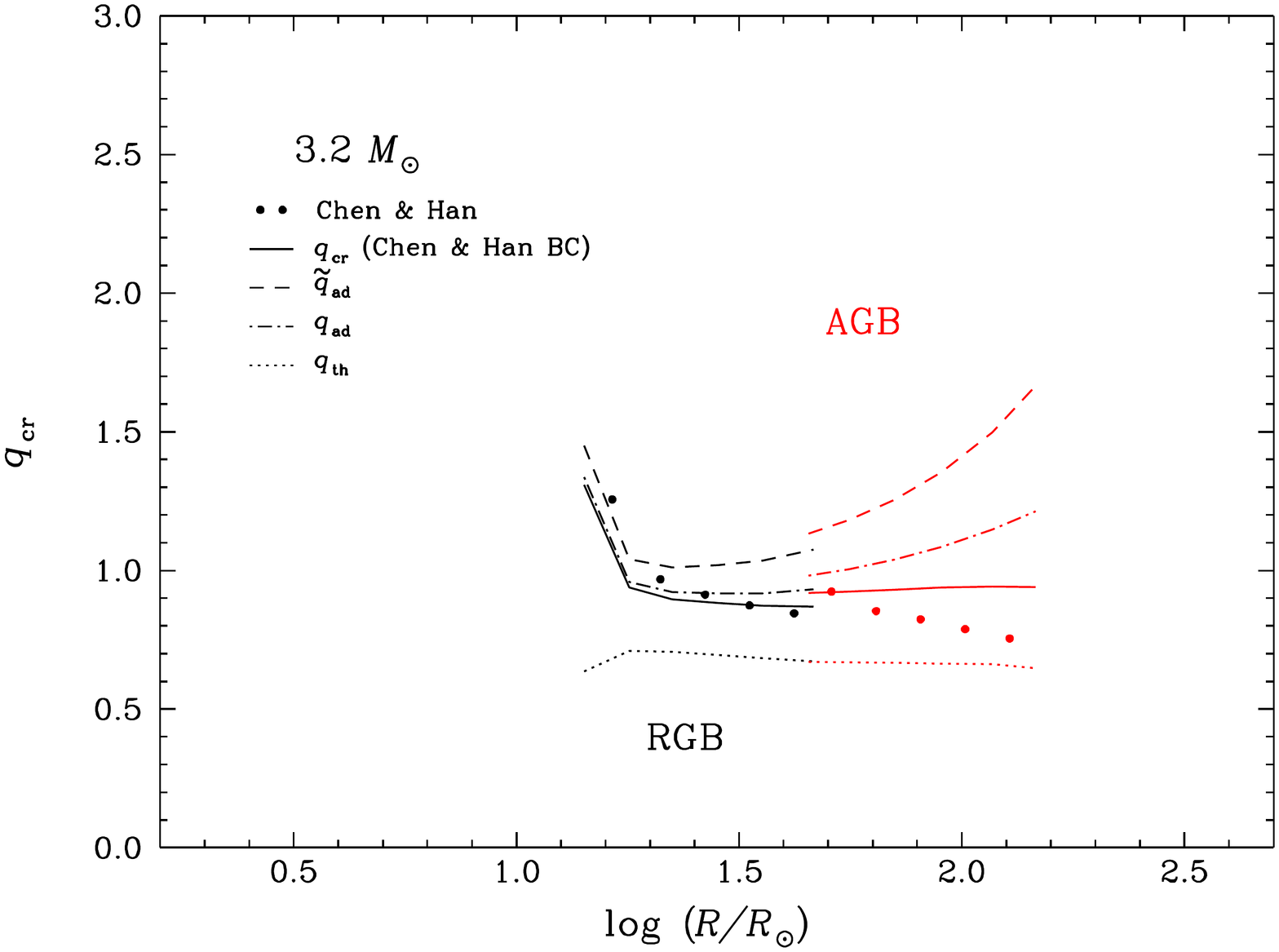}
\caption{Critical mass ratios, $q_{\rm cr}$, for the onset of dynamical timescale mass transfer derived from different calculations. $q$ derived from the time-dependent calculation in \citealt{chen08}, $q_{\rm cr}$ adiabatic mass loss calculation with same boundary condition as \citeauthor{chen08}, $q_{\rm ad}$ adiabatic mass loss calculation from stars with a standard mixing-length envelope, and $\tilde{q}_{\rm ad}$ adiabatic mass loss calculation from stars with an isentropic envelope are present as dots, solid line, dash-dotted line, dashed line, respectively. $q_{\rm th}$ for the onset of thermal timescale mass transfer derived from thermal equilibrium mass loss model are also shown with thin dotted line. Black lines and red lines indicate the value of $q_{\rm cr}$ from RGB and AGB stellar models, respectively. \label{3_2MsunBC}}
\end{figure}

\begin{figure}
	\centering
	\includegraphics[scale=0.48]{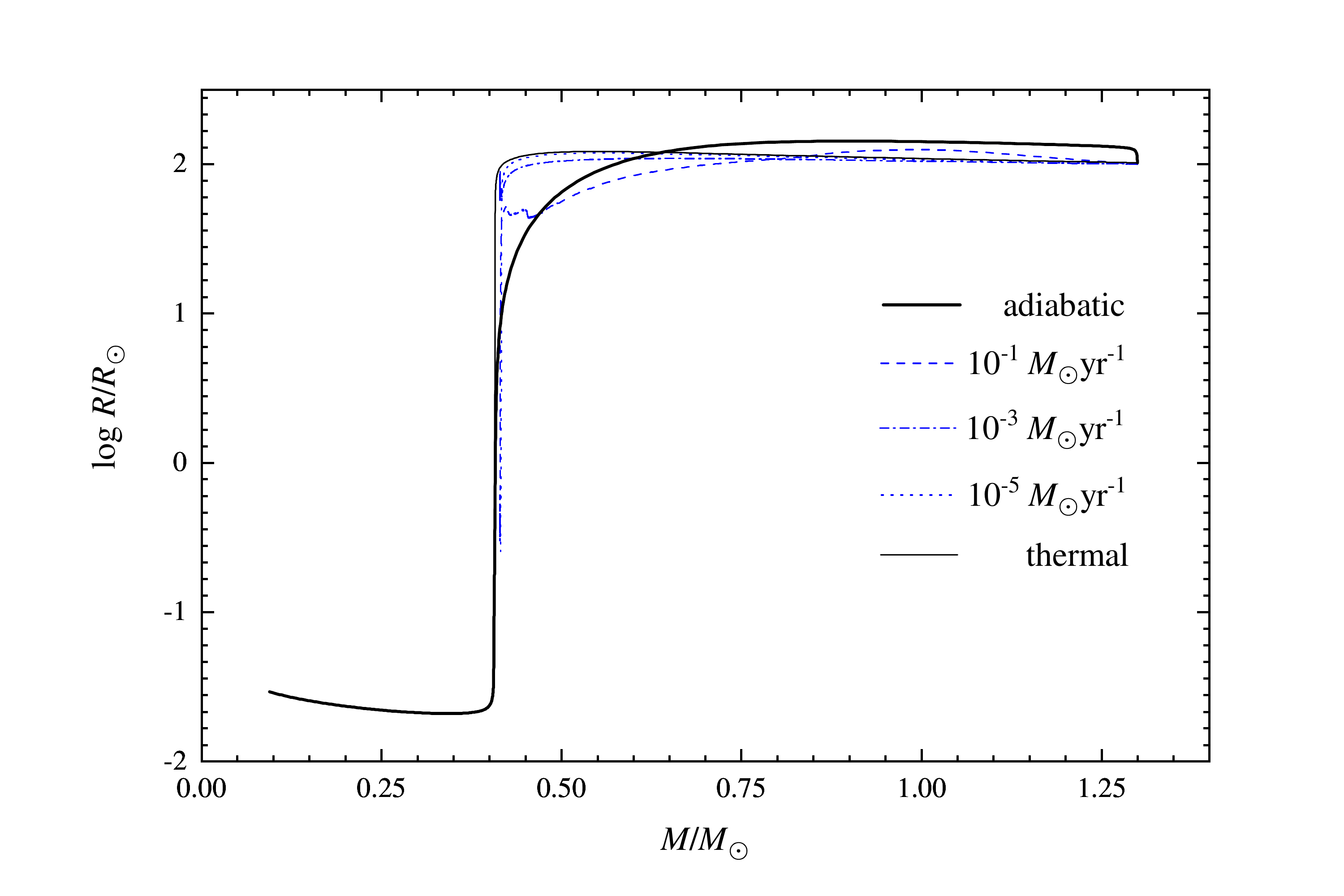}
	\caption{Radius response to different mass-loss rate, adiabatic, and thermal equilibrium mass loss of a $1.3 M_\sun$ red giant branch (RGB) star. The thick solid line and thin solid line show the radius response to adiabatic mass loss and thermal equilibrium mass loss, respectively. Blue lines represent how the radius responds to time-dependent calculations (from MESA) with constant mass-loss rates of $1\times10^{-5}$, $1\times10^{-3}$, and $1\times10^{-1} M_\sun {\rm yr}^{-1}$. For this $1.3 M_\sun$ RGB star,the stellar radius response to rapid mass loss lies between adiabatic and thermal equilibrium mass loss. \label{1_3Msun-RM}}
\end{figure}

\begin{figure}
	\centering
	\includegraphics[scale=0.48]{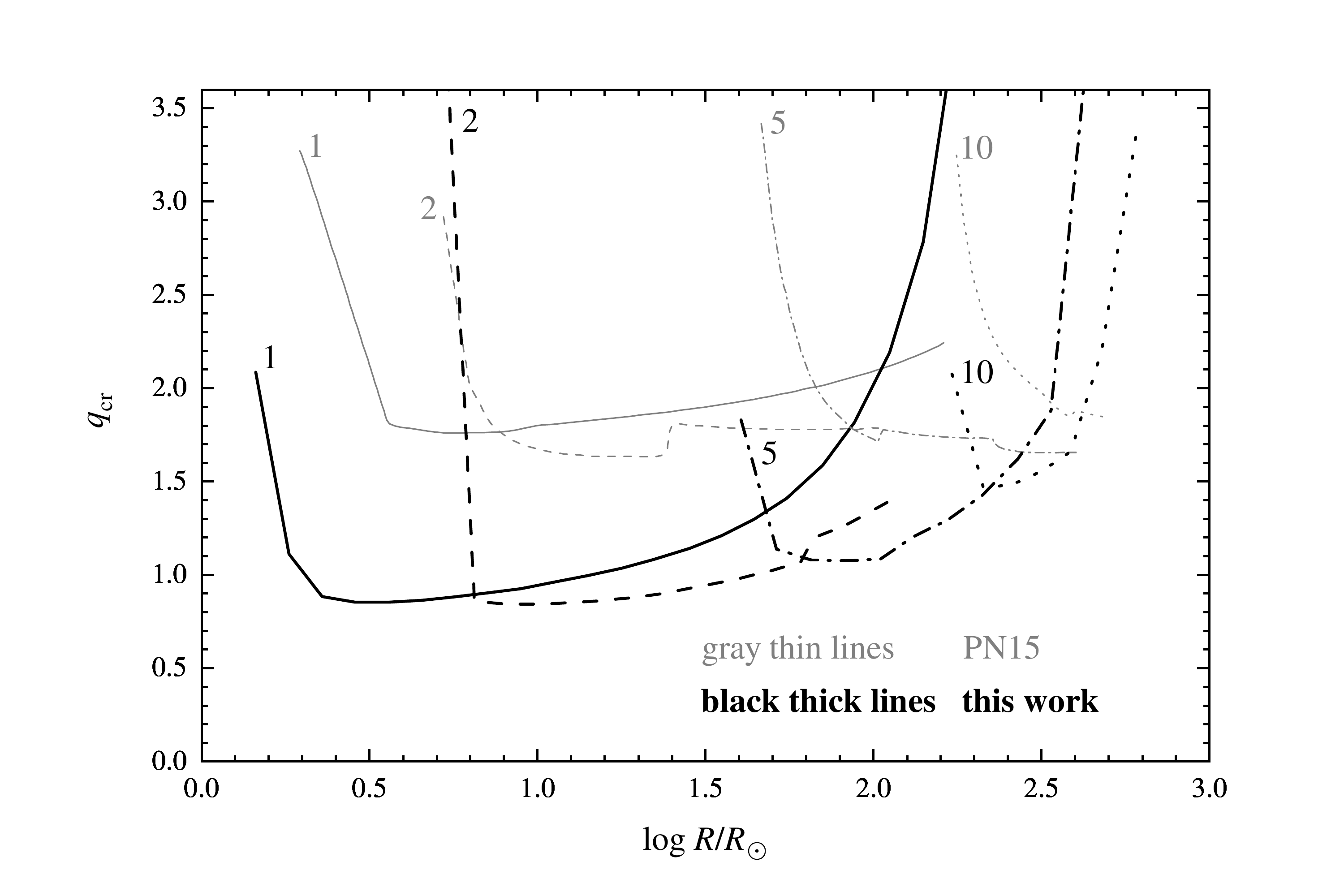}
	\caption{Critical mass ratios, $q_{\rm cr}$, for dynamical timescale mass transfer of the giant branch stars. Solid, dashed, dashed-dotted, and dotted lines indicates the $q_{\rm cr}$ of the $1,~2,~5,~{\rm and}~10~M_\sun$ giant branch stars, respectively. Black thick lines are from this work and gray thin lines come from \cite{pavl15}.  \label{1_3Msun-q}}
\end{figure}

\begin{figure}
\centering
\includegraphics[scale=0.6]{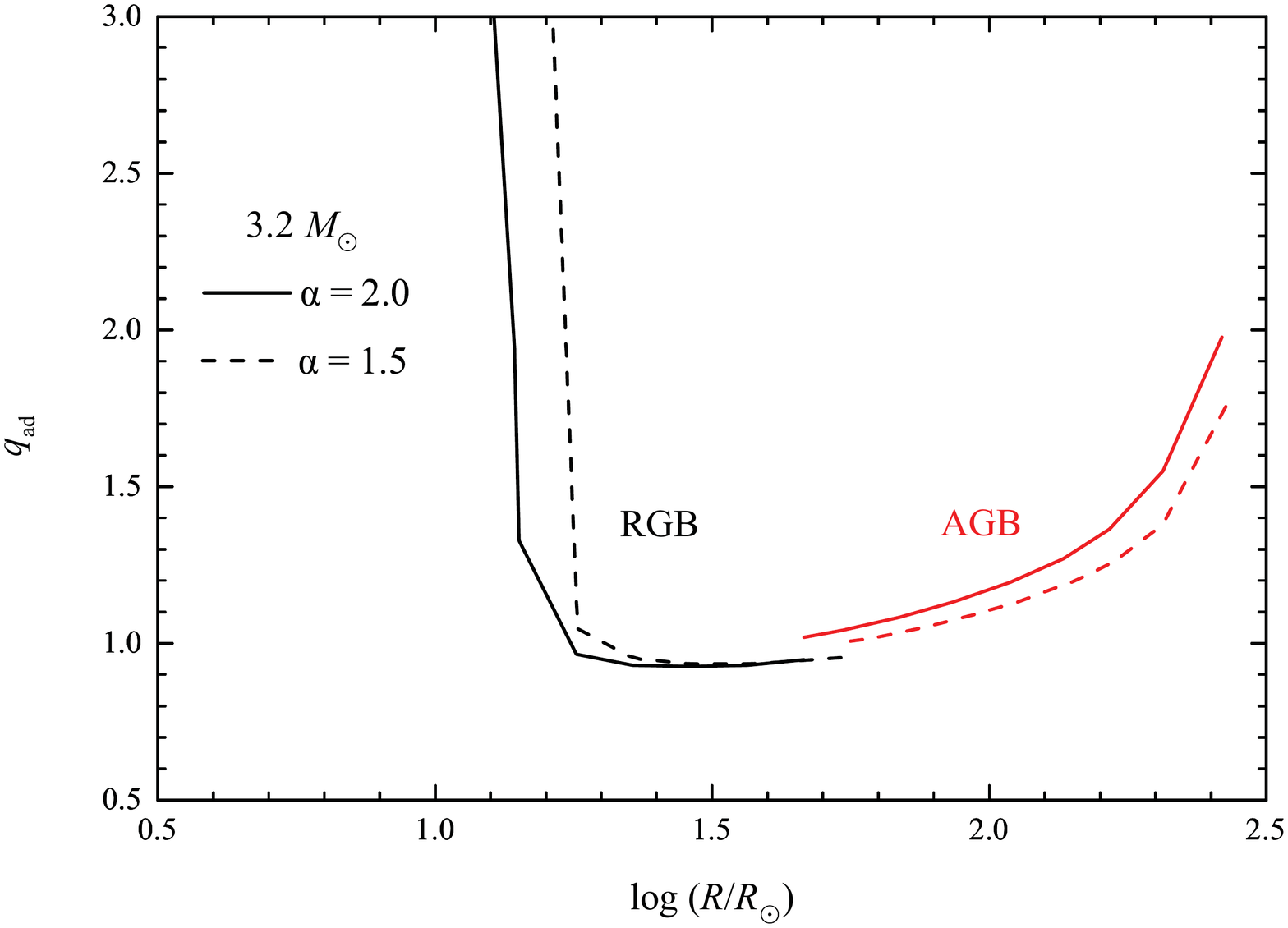}
\caption{Critical mass ratios, $q_{\rm ad}$, for the onset of dynamical timescale mass transfer derived from adiabatic mass loss calculation of RGB (black lines) and AGB (red lines) stars with different mixing length. \label{3_2Msun_alpha}}
\end{figure}

When we compare the critical mass ratio $q_{\rm cr}$ between our results and those derived from time-dependent calculations, we find that a realistic model of mass transfer rate as a function of Roche lobe overfill is important. The critical mass ratio for dynamical instability is sensitive to how we assume the formula of mass transfer rate. Normally, people use the formula of mass transfer rate similar to that in \citet{chen08} or in \citet{wood11}. This kind of formula physically assumes the mass that overfills inner Lagrangian point would be transfered out of it immediately. However, the mass transfer might be limited by the local sound speed. What we find is that the density at the extended envelope of RGB and AGB stars is very low, the mass transfer rate is hard to increase to a very high value. Apart from this, as we discussed in Section 2 and 3, we use the interior radius $R_{\rm KH}$ instead of its surface radius $R$ to quantify the mass-radius exponent. If we set the boundary condition for mass transfer rate $\dot{M}$ (formula A9 in Paper I) the same as in \citet{chen08}, we get the similar pattern of $q_{\rm ad}$, which decreases with elder evolutionary stages on RGB and AGB (dots in Figure~\ref{3_2MsunBC}). On the other hand, we get the opposite pattern of $q_{\rm ad}$, which increase with later evolutionary stages on RGB and AGB (dashed and dash-dotted lines for standard mixing length envelope and isentropic envelope stellar models, respectively, in Figure~\ref{3_2MsunBC}). This is due to that the RGB and AGB stars become harder to reach a thermal timescale as the thermal timescale getting shorter and the extended low-density envelope getting bigger on an later evolutionary stage.

In addition to the comparison above, we also compared the radius response from our model to that of mass loss calculations of MESA. We take a $1.3~M_\sun$ RGB donor star ($\log (R/R_\sun) = 2.0$) as an example. We set the mass loss rates to be $1\times10^{-5}$, $1\times10^{-3}$, and $1\times10^{-1} M_\sun {\rm yr}^{-1}$, respectively, and use the MESA code of version 12115 \citep[e.g.,][]{paxt19} to do the calculations. Figure~\ref{1_3Msun-RM} shows the radius responses in the calculations of MESA and the response from our adiabatic/thermal equilibrium mass loss model \citep{ge15,ge20}. We find that a big mass loss rate leads to a radius response closer to that from our adiabatic mass loss model. There also exists a trend that the radius response is closer to that from our adiabatic mass loss model at the first half of mass loss except for the surface superadiabatic expansion (see Section~\ref{sec_supad}), and is then closer to that from our thermal mass loss model when the core is nearly naked. This is due to that the thermal time scale is much shorter for a thin envelope. The radius response of the RGB donor star to rapid mass transfer is nicely constrained by a lower limit from the thermal equilibrium mass loss model and the upper limit from the adiabatic mass loss model. Recently, \citet{pavl15} used the MESA code to study the response of the giant branch donor star to rapid mass loss. Combined with the polytope model, the L2/3-overflow, and the radius response from the MESA code, they derived the critical mass ratio for dynamical timescale mass transfer (see the thin lines in Figure~\ref{1_3Msun-q}). We compared their critical mass ratios to our results. We find that the critical mass ratios of the $1,~2,~5,~{\rm and}~10~M_\sun$ giant branch stars from our results have the same pattern with those from their results. The critical mass ratio decreases sharply in the "transition" region (see Figuree~\ref{3_2zeta}) and then slightly increases with the growing of the convective envelope. For the late RGB/TPAGB stars, the thermal timescale is already very short (about $10^2$ yr). Therefore, these donor stars with an extended envelope are very hard to reach a dynamical timescale mass transfer and the critical mass ratio of these donors tends to be very high in our calculation. However, there is a systematic difference between our results and their results. We suspect that part of the difference is caused by the assumption of their "L2/L3-overflow condensed polytrope simplification". Because that we derive the critical mass ratio by comparing the response of the interior radius and the inner Lagrangian radius to adiabatic mass loss. And the outer Lagrangian radius is slightly larger than the inner Lagrangian radius.

We also test the sensitivity of our results to the choice of different mixing-length in the convection envelope. Taking $3.2 M_\odot$ stars as an example here in Figure~\ref{3_2Msun_alpha}, we find that as the mixing-length parameter decreases from 2 to 1.5, RGB maximum radius and AGB minimum radius have a rightward shift. However, the critical mass ratio, $q_{\rm ad}$, still displays the same pattern at different evolutionary stages (as a function of radius $R$). During this process, the values of $q_{\rm ad}$ as a function of $R$ varies monotonically, as does the RGB maximum radius and AGB minimum radius. Which means, if we move the solid lines rightward, the dashed lines would coincide with the solid lines. So we would argue that the mixing-length parameter would affect the stellar structure, but it does not play an important role in the value of $q_{\rm ad}$.

\section{DISCUSSION and CONCLUSIONS}
\label{dis_con}

This study is the third paper that attempts for the first time to survey the thresholds systematically for dynamical timescale mass transfer over the entire span of possible donor star evolutionary states. These thresholds mark bifurcation points in close binary evolution, separating evolutionary channels proceeding on a thermal timescale (or slower) from those proceeding on a far more rapid timescale leading to common envelope evolution. We can apply these thresholds as input to population synthesis studies of close binary evolution that seek to quantify the frequency and properties of various possible evolutionary channels. We survey the adiabatic responses of Population I ($Z = 0.02$) stars spanning a full range of stellar mass ($0.10\ M_{\odot}$ to $100\ M_{\odot}$) and evolutionary stages from the ZAMS through the MS, HG, RGB, and AGB, up to the exhaustion of the hydrogen-rich envelopes, carbon ignition, or core-collapse, as the case may be.

We assume the donor response to rapid mass loss is one of adiabatic expansion throughout the donor interior to its surface. This approximation may be valid throughout the bulk of the interior once the mass loss rate significantly surpasses the thermal timescale rate, but it may break down near the stellar photosphere, where radiative relaxation becomes extremely rapid. More significantly, we need to notice that the growth to supra-thermal mass transfer rates generally extends over many thermal timescales, while we approximate the donor response even in these circumstances as being purely adiabatic. An estimate of the amount of mass lost during the acceleration to dynamical timescale can be had from the difference $M - M_{\rm KH}$ in Table~\ref{dynml}. For RGB and AGB stars with moderately deep surface convection zones, thermal relaxation during this acceleration phase is probably of little consequence because convection zones tend to respond as coherent entities (specific entropy rises or falls more or less uniformly throughout the convection zone), but more significantly, because such stars are subject to prompt dynamical instabilities that cut short the acceleration phase. For donor stars on the MS and HG with a radiative envelope, what we find in Paper II is that a vast majority of binaries with mass ratios exceeding $\tilde{q}_{\rm ad}$ for the delayed dynamical instability will, in fact, evolve into contact before actually reaching the point of instability if they have non-degenerate accretors. For donor stars on the RGB and AGB with a deep enough convective envelope, we also find that binaries containing these kinds of donor stars may evolve into a common envelope phase even when donor stars are dynamically stable. This can be explained if the thermal timescale of these donor stars is short enough, and the rapid mass transfer process caused the donor star to overfill its Roche lobe by a by a significant margin (10-20 percent of its radius in some cases), even overfilling its outer Lagrangian point. If the donor star is transferring its mass to a degenerate companion, the latter may not absorb the transferred mass quickly enough, allowing an extended structure to form. In this case, both of the companions will overfill their Roche lobes, and a common envelope may form even on thermal timescales. This kind of situation raises the need for thermal-equilibrium mass loss models, which can be used to study the response of donor stars suffering a thermal timescale mass transfer. We have established the thermal equilibrium mass loss model in a separate study. Some of results of that study are shown in Paper II and this paper. We will present the complete set of results in the next series of papers \citep{ge20}.

\begin{figure}
\centering
\includegraphics[scale=0.4]{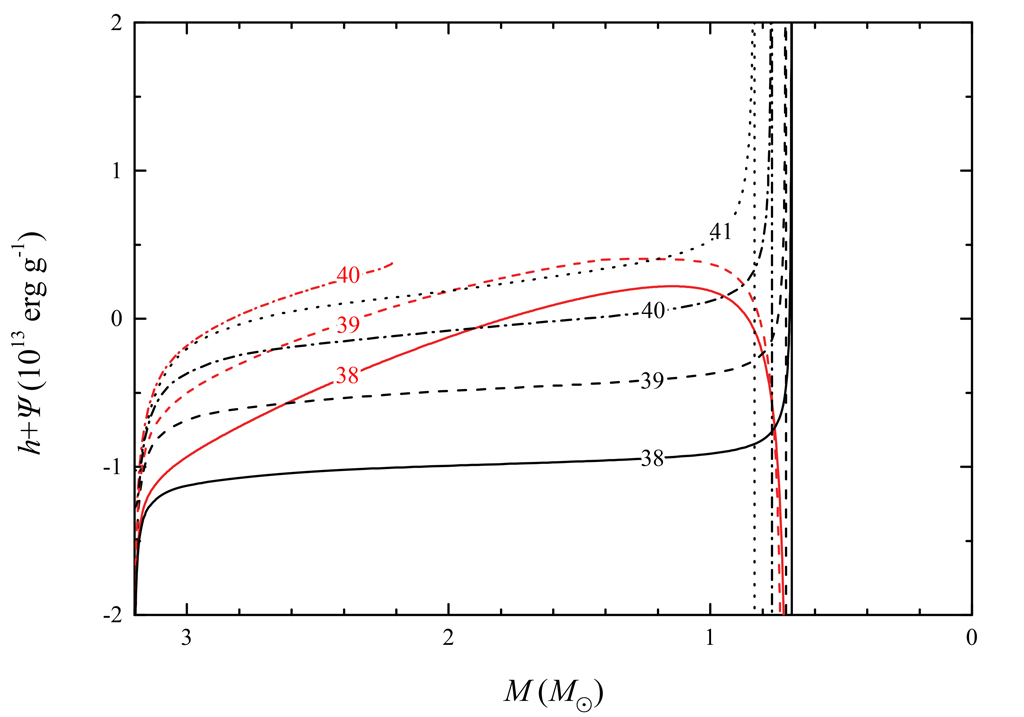}
\caption{$h + \Psi$ profile as a function of mass (black lines), and $h + \Psi$ on the surface of the star as a function of remaining mass (red lines) are presented in this figure. $3.2 M_\odot$ AGB stars with stellar model number $k= 38 , 39, 40, 41$ in Table~\ref{intmod},~\ref{glbmod}, and~\ref{dynml} are shown by solid, dashed, dash-dotted, dotted lines, respectively. Based on the Bernoulli equation, red lines indicate that it is energetically possible for the stellar surface layers to escape to infinity. Stellar model number $k=41$ terminate unexpectedly for a positive and higher value of $h + \Psi$. \label{3_2Msun_unbound}}
\end{figure}

We briefly discuss the physical basis for failure of the most luminous AGB models (TPAGB stars) to converge (gray zone in Figure~\ref{qcrit_real} and Figure~\ref{qcrit_isen}). We examine the question of whether removal of the superadiabatic surface layer can allow the underlying stellar envelope to expand freely to infinity. If we consider an adiabatic streamline originating at some point in the stellar envelope, then for inviscid flow the Bernoulli equation tells us that ($v^2/2 + h + \Psi$) is a conserved quantity, where $v$ is the local velocity of that streamline, $h$ the specific enthalpy of the gas, and $\Psi$ the gravitational potential energy per unit mass at the point in question. At the origin of a streamline, the initial velocity in the static envelope is of course $v = 0$, so if ($h + \Psi$) at that point is positive, it is at least energetically possible for that streamline to extend to infinity ($\Psi -> 0$, with $v^2 -> 0$). We take $3.2 M_\odot$ AGB stars as examples. In Figure~\ref{3_2Msun_unbound}, the black curves give the ($h + \Psi$) profile as a function of mass through the envelope of the initial model (models $k= 38 , 39, 40, 41$ in Tables~\ref{intmod},~\ref{glbmod}, and~\ref{dynml} are shown by solid, dashed, dash-dotted, dotted lines, respectively) in a mass loss sequence, with the surface of the star to the left. Mass is removed from the surface of the star, which relaxes adiabatically to hydrostatic equilibrium. The red curves provide the values of ($h + \Psi$) at the surface of the star as a function of remaining mass along a mass loss sequence (models $k= 38, 39, 40$ as in Tables~\ref{intmod},~\ref{glbmod}, and~\ref{dynml} are shown by solid, dashed, dash-dotted, respectively). We see that it is indeed energetically possible (though not necessarily so) for the stellar surface layers to escape to infinity. We are reasonably certain that this implicit breakdown of hydrostatic equilibrium at the surface of the donor star lies at the root of the failure of our AGB mass loss sequences to converge (gray zone in Figure~\ref{qcrit_real} and Figure~\ref{qcrit_isen}).

\begin{deluxetable}{rrrrrrrr}    
	\tabletypesize{\footnotesize}
	\tablewidth{0pt}
	\tablecolumns{8}
	\tablecaption{Comparison of $q_{\rm cr}$ for stellar models with and without winds\label{wind}}
	
	\tablehead{
		\colhead{${M_{\rm ZAMS}}$} & \colhead{$M_{\rm RLOF}$} & \colhead{$M_{\rm core}$} & \colhead{$\log R$} &
		\colhead{$q_{\rm cr} = M_{\rm RLOF}/M_2$} & \colhead{$\psi_{\rm c}$} & \colhead{${\rm core~type}$} &
		\colhead{${\rm Note}$} \\
		\colhead{$M_\sun$} & \colhead{$M_\sun$} & \colhead{$M_\sun$} & \colhead{$R_\sun$} &
		\colhead{} & \colhead{${\rm k}T$} & \colhead{} & \colhead{} 
	}
	\startdata
	\cutinhead{\normalsize low-mass stars}
	1.60&	1.30&	0.406&	1.966&	1.36&	17.84&	deg He core&    wind 1 \\*
	1.60&	1.60&	0.406&	1.958&	1.33&	17.62&	deg He core&	no-wind \\
	1.60&	1.60&	0.408&	1.966&	1.34&	17.65&	deg He core&	no-wind \\
	1.30&	1.30&	0.406&	2.010&	1.61&	18.57&	deg He core&	no-wind \\
	1.30&	1.30&	0.395&	1.966&	1.52&	18.53&	deg He core&	no-wind \\
	\cutinhead{\normalsize massive stars}
	16.00&	13.00&	3.581&	2.790&	4.11&	-3.74&	non-deg CO core&    wind 2 \\
	16.00&	16.00&	3.581&	2.879&	4.14&	-3.78&	non-deg CO core&	no-wind \\
	16.00&	16.00&	1.771&	2.790&	2.91&	-3.51&	non-deg CO core&	no-wind \\
	13.00&	13.00&	2.752&	2.790&	2.71&	-1.73&	non-deg CO core&	no-wind \\
	\enddata
	\tablecomments{
	 1. $M_{\rm ZAMS}$ is the stellar mass on the zero-age main-sequence (ZAMS); \\
	 2. $M_{\rm RLOF}$ is the mass of the donor star at the onset of Roche-lobe overflow (RLOF);\\
	 3. $M_{\rm core}$ is the core mass of the donor star at the onset of RLOF;\\
	 4. $\log R$ is the radius of the donor star at the onset of RLOF; \\
	 5. $M_2$ is the mass of the companion at the onset of RLOF; \\
	 6. $q_{\rm cr}$ is the initial critical mass ratio for dynamical timescale mass transfer; \\
	 7. $\psi_{\rm c}$ is the degeneracy parameter in the core of the donor at the onset of RLOF; \\
	 8. wind 1 -- we adopt the Reimers' wind by \citet{reim75}. For the convenience of the comparison with our model grid, we set the free parameter $\eta=3$; \\
     9. wind 2 -- we adopt the wind formula for massive stars from \citet{vink01} and \citet{jage88} for temperature larger or less than 15000 K, respectively; \\
     10. deg is the short abbreviation of degenerate.}
\end{deluxetable}
\newcounter{tbl4}

Stellar winds may significantly change the masses of both the donor and the accretor as well as the orbital separation before Roche-lobe filling. They are usually taken into account in many binary evolution calculations and binary population synthesis studies. However, our initial donor stars that fill their Roche-lobes are built without addressing any wind mass-loss. We therefore checked the effect of stellar winds on the critical mass ratio for a low mass star and a massive star. For the case of the low mass star, we choose a $1.6~M_\sun$ ZAMS star and evolve it to RGB with Reimers’ wind \citep{reim75}. Reimers coefficient $\eta=1/3$ is widely adopted. However, we set the coefficient $\eta=3$ for the convenience of comparison with our model grid, and doing so does not affect our conclusion. When the mass of the star becomes $1.3~M_\sun$ due to wind mass loss, we get a model with a core mass of $0.406~M_\sun$ and a radius of $\log (R/R_\sun)=1.966$. Using the adiabatic mass loss code, we calculate the critical mass ratio of this $1.3~M_\sun$ RGB star. We then compare the critical mass ratio from this model to that from $1.6~M_\sun$ and $1.3~M_\sun$ stellar models (without wind) with the same core mass or the same radius. The detailed results are listed in Table~\ref{wind}. For the case of the massive star, we choose a $16.0~M_\sun$ ZAMS star and evolve it to the giant branch with a widely used wind (\citealt{vink01} for $T>15000{\rm K}$; \citealt{jage88} for $T<15000{\rm K}$). When the mass of the star becomes $13.0~M_\sun$, we get a model with a core mass of $3.581M_\sun$ and a radius of $\log R/R_\sun =2.790$. Similar to the case of the low mass star, we calculate the critical mass ratio of this $13.0~M_\sun$ giant branch star. We then compare the critical mass ratio from this model to that from $16.0~M_\sun$ and $13.0~M_\sun$ giant branch stars (no wind) with the same core mass or the same radius. The detailed results are listed in Table~\ref{wind}. Note that the maximum core mass of $13.0~M_\sun$ giant branch stars without wind could not reach to $3.581M_\sun$. Therefore, it seems like that one line in the lower part is missed compared with the upper part of the Table. We compare the critical mass ratios derived from stellar models with wind and the models without wind. We find that the critical mass ratio of a giant star with wind and with a core mass Mcore is very close to that of a giant star without wind but with the same ZAMS mass and the same core mass. This might be caused by the fact that, for donor stars on the giant branch, cores dominate the fundamental structure and their corresponding envelope’s response. We therefore suggest to take the critical mass ratio of a giant star (with an initial mass $M_{\rm ZAMS}$ and a core mass $M_{\rm core}$) with wind to be that of a star in our model grid (without wind) with the same ZAMS mass $M_{\rm ZAMS}$ and the same core mass $M_{\rm core}$.

As we discussed in Section \ref{comparison}, our results show that RGB and AGB stars tend to be more stable than previously believed (\cite{pavl15} also got similar results for giant branch stars with $q_{\rm cr}$ around 1.3 to 2.7). A larger critical mass ratio may be helpful to explain the abundance of observed post-AGB stars with an orbital period of around 1000 days. Given a prescription for how the donor Roche lobe responds to mass loss, our adiabatic mass-loss sequences are, in principle, applicable to nonconservative mass transfer as well (more detailed discussion can be found in Paper II). Adiabatic mass-loss sequences can also be used to constrain the survival of binaries entering into common envelope evolution through dynamical timescale mass transfer. With the total energy that is a function of remaining mass in the adiabatic mass loss process, we can combine energy constraints with the requirements that both binary components fit within their post-common envelope Roche lobes. Thus, we can place strict limits on the mass, mass ratios, and remnant orbital separations of binaries passing through common envelope evolution. We will present our results in the next paper of this series.

\acknowledgments

We are grateful to the anonymous referee for the helpful comments and suggested improvements.This work was supported by grants from the National Natural Science Foundation of China (NSFC Nos. 11673058, 11733008), the key research program of frontier sciences, CAS, No. ZDBS-LY-7005, the Natural Science Foundation of Yunnan Province (Grant No. 2019HA012), and the Department of Astronomy at the University of Illinois at Urbana-Champaign. We thank Yan Gao for proof reading this paper and partial language support. HG thanks the Chinese Academy of Sciences and the Department of Astronomy, the University of Illinois at Urbana-Champaign, for the one-year visiting program. HG also thanks Dr. Hailiang Chen for providing the time-dependent mass loss calculations from MESA. RFW was supported in part by US National Science Foundation grants AST 04-06726 and AST 14-13367.

\end{document}